\documentclass[
 reprint,
 superscriptaddress,
 amsmath,amssymb,
 aps, nofootinbib
]{revtex4-2}

\usepackage{graphicx}
\usepackage{dcolumn}
\usepackage{bm}
\usepackage{acronym}
\usepackage{booktabs}
\usepackage{float}
\usepackage{pifont}
\usepackage{siunitx}
\usepackage{placeins}
\usepackage{mathrsfs}
\usepackage[utf8]{inputenc}
\usepackage[normalem]{ulem}
\usepackage[breaklinks,colorlinks,citecolor=blue]{hyperref}
\usepackage{graphicx,psfrag}
\usepackage{placeins}
\usepackage{multirow}
\usepackage{comment}
\usepackage{ulem}
\usepackage{hyperref}
\usepackage{color}
\usepackage{xcolor}
\usepackage{comment}
\usepackage{natbib}

\begin{document}


\title{General relativistic hydrodynamic simulations of binary strange star mergers}

\author{Francesco Grippa}
\affiliation{Dipartimento di Fisica ``E.R. Caianiello'', Universit\`{a} degli Studi di Salerno, Via Giovanni Paolo II, 132, 84084 Fisciano (SA), Italy}
\affiliation{INFN, Sezione di Napoli, Gruppo
Collegato di Salerno, Via Giovanni Paolo II, 132, 84084 Fisciano (SA), Italy}

\author{Aviral Prakash}
\affiliation{Institute for Gravitation \& the Cosmos, The Pennsylvania State University, University Park PA 16802, USA}
\affiliation{Department of Physics, The Pennsylvania State University, University Park PA 16802, USA}
\affiliation{Department of Physics, University of California, Berkeley CA 94720, USA}
\affiliation{Department of Physics and Astronomy, University of New Hampshire, Durham NH 03824, USA}

\author{Domenico Logoteta}
\affiliation{Dipartimento di Fisica, Universit\`{a} di Pisa, Largo B.  Pontecorvo, 3 I-56127 Pisa, Italy}
\affiliation{INFN, Sezione di Pisa, Largo B. Pontecorvo, 3 I-56127 Pisa, Italy}

\author{David Radice}
\thanks{Alfred P.~Sloan Fellow}
\affiliation{Institute for Gravitation \& the Cosmos, The Pennsylvania State University, University Park PA 16802, USA}
\affiliation{Department of Physics, The Pennsylvania State University, University Park PA 16802, USA}
\affiliation{Department of Astronomy \& Astrophysics,  The Pennsylvania State University, University Park PA 16802, USA}

\author{Ignazio Bombaci}
\affiliation{Dipartimento di Fisica, Universit\`{a} di Pisa, Largo B.  Pontecorvo, 3 I-56127 Pisa, Italy}
\affiliation{INFN, Sezione di Pisa, Largo B. Pontecorvo, 3 I-56127 Pisa, Italy}

\begin{abstract}
 We perform fully general-relativistic simulations of binary strange star mergers considering two different approaches for thermal effects. The first uses a cold equation of state (EOS) derived from a modified version of the MIT bag model which is then supplemented by a $\Gamma$-law correction. The second approach employs a microphysical description of the finite-temperature effects. We describe the results obtained with the two treatments, highlighting the influence of thermal effects. We find that the postmerger dynamics differs significantly in the two cases, leading to quantitative differences in the postmerger gravitational-wave spectrum and ejecta mass. The peak frequency of the postmerger gravitational-wave emission is consistent with the established quasi-universal relations for binary neutron star mergers and as a result, our simulations cannot distinguish between mergers of neutron stars and those of strange stars. Our models with realistic treatment of finite-temperature effects produce a significant amount of ejecta $\gtrsim 0.02\ M_{\odot}$. The resulting flux of strangelets near the Earth, computed assuming that all neutron star mergers are in fact strange-stars mergers, that the average mass of each strangelet is $A \sim 100$, and that the binary considered here is representative, is in tension with experimental upper limits.
\end{abstract}

\maketitle

\section{Introduction}
\label{sec_Introduction}

The detection of gravitational waves (GWs) from the coalescence of the binary neutron star (BNS) merger GW170817 \cite{TheLIGOScientific:2017qsa} complemented by the subsequent observations of electromagnetic (EM) counterparts by a number of Earth and space-based telescopes \cite{GBM:2017lvd}, has opened a new avenue for exploring the properties of matter under extreme densities (up to several times the nuclear saturation density $\rho_{\rm nuc} \sim 2.7 \times 10^{14} \; \rm{g~ cm^{-3}}$) and temperatures (up to $10$\rm{s} of \rm{MeVs}) that cannot be presently realized in any terrestrial laboratory. These extreme physical conditions are in fact expected to occur in the post-merger remnants generated in BNS mergers and in core-collapse supernovae \cite{Bombaci_NPA_1995, M.Prakash:PhysRep_1997, Perego:2019adq}. The equation of state (EOS) describing the thermodynamic properties of this extreme matter primarily determines the evolution, the fate of the merger remnant, the GW signal, and to some extent, the properties of the ejecta and associated electromagnetic counterparts. Other effects such as neutrino transport, \cite{Radice:2016dwd, Radice:2018pdn, Radice:2021jtw, Schianchi:2023uky, Foucart:2022bth, Radice:2023zlw, Zappa:2022rpd, Loffredo:2022prq, Camilletti:2022jms, Most:2021zvc, Most:2022yhe, Combi:2022nhg, George:2020veu, Siegel:2017jug, Martin:2017dhc, Fujibayashi:2020qda, Grohs:2022fyq, Grohs:2023pgq, Richers:2019grc}, and magnetohydrodynamic turbulence \cite{Ciolfi:2020cpf, Ciolfi:2017uak, Radice:2017zta, Shibata:2017xht} may also influence the post-merger evolution. 

In the aforementioned hot and dense region of the phase diagram of strongly interacting matter, quantum chromodynamics (QCD) predicts a transition from a regime where quarks and gluons are confined within baryons and mesons (the hadronic matter phase) to a regime with deconfined quarks and gluons: the so-called quark matter phase. In this region of the QCD phase diagram, it is still an open question as to whether the hadronic-to-quark matter phase transition is of the first order or proceeds continuously as in a crossover \cite{Baym:2017whm}. 
Upcoming experiments such as the Compressed Baryonic Matter (CBM) experiment \cite{senger2021} at the Facility for Antiproton and Ion Research (FAIR) 
are expected to shed light on the nature of such a phase transition. 

It has long been suggested that a phase of matter made up entirely of the three lightest quark flavors, the up ($u$), down ($d$), and strange ($s$) quarks along with a sufficient number of electrons to ensure electrical neutrality, could exist in the core of sufficiently massive neutron stars (NS).  
This form of quark matter is referred to as strange quark matter (SQM) and a NS possessing a SQM core is known as a hybrid star (see e.g., \cite{1997csnp.book.....G}). 
Even more intriguing than the existence of an SQM core in a NS is the possible existence of a new family of compact stars, called strange stars, 
which are entirely (i.e., up to the surface) made up of SQM (bare strange stars). They could possibly be covered by a thin crust of ``normal'' matter similar to the one found in the outer crust of a NS (strange stars with crust). 
The possible existence of strange stars is a consequence of the so-called Bodmer--Witten hypothesis \cite{Bodmer:1971we,Witten:1984rs}. 
According to this hypothesis, SQM is absolutely stable, i.e., its energy per baryon $(E/A)_{uds}$ (at the baryon density where the pressure is equal to zero) 
is less than the energy per baryon of the most bound atomic nuclei ($^{56}\rm{Fe}, ^{58}\rm{Fe}, ^{62}\rm{Ni}$) which is $\sim 930.4~\rm{MeV}$.  
The absolute stability of SQM does not preclude the existence of ``ordinary'' matter \cite{Bodmer:1971we,Witten:1984rs}.   
In fact, under this hypothesis, atomic nuclei can be considered as metastable states (with respect to the decay to SQM droplets) having a mean-life time many orders of magnitude larger than the age of the Universe. 

Over several years, a number of compact stars associated with different astrophysical phenomena have been proposed as possible candidates for strange stars 
\cite{Dey+1998,SaxJ1808,XDLi_1999,Drake+2002,Weber:2004kj}.  
Most recently, it has been argued that the low-mass companion (with a mass in the range $(2.50 - 2.67)$  $M_\odot$) of the $23~M_\odot$ black hole, whose merger generated the gravitational wave signal GW190814 \cite{GW190814:2020}, could be a strange star \cite{Bombaci:2020vgw}. Massive strange stars could thus populate the so-called mass-gap between neutron stars (NSs) and black holes (BHs). 
It has also been suggested that the Central Compact Object within the supernova remnant HESS J1731-347 \cite{ WOS:000871277600001} could be interpreted as a low-mass 
($M = 0.77^{+0.20}_{-0.17} M_\odot$) strange star \cite{WOS:000871277600001,DiClemente:2022wqp,Horvath:2023uwl}. 

Strange stars could be formed in supernova explosions \cite{Benvenuto-Horvath_1989,Fischer:2010wp}, during the early evolution of a protoneutron star \cite{Mintz:2009ay,Bombaci:2009jt,Bombaci:2011mx,Lugones:2015bya}, or by an external seeding of strangelets (chunks of SQM) \cite{Farhi:1984qu} in ordinary neutron stars \cite{Olinto:1986,Bucciantini:2019ivq}. 
Alternatively, the conversion of an ordinary neutron star into a strange star is a very likely formation mechanism. Works like \cite{Herzog:2011sn, Pagliara:2013tza} have studied the dynamics of such a conversion process assuming the presence of an initial SQM seed and modeling the conversion as a turbulent deflagration. In fact, as it has been proposed and discussed in several works \cite{Berezhiani:2002ks,Bombaci:2004mt,Drago:2004vu}, if the hadronic-to-quark matter phase transition is of the first order, ordinary NSs above a threshold value of their  gravitational mass (corresponding to a threshold central density $n_{thr}$) become metastable with respect to the conversion to strange stars. These metastable NSs have a {\it mean-life time} related to the nucleation time $\tau$ to form the first critical-size SQM droplet in their center\footnote{The actual mean-life time of the metastable NS depends on the mass accretion or on the spin-down rate which modifies the nucleation time via an explicit time dependence of the stellar central density $n_c$.}. As shown in \cite{Berezhiani:2002ks,Bombaci:2004mt,Drago:2004vu}, $\tau$ decreases very steeply as a function of the stellar central density $n_c$ (or as a function of the corresponding gravitational mass $M_G(n_c)$), from $\tau = \infty$ when $n_c = n_{thr}$, to values much smaller than typical pulsar ages (see e.g. Fig. 1 in \cite{Berezhiani:2002ks}). At this point (e.g. when $\tau \sim 1 \,$yr \cite{Berezhiani:2002ks,Bombaci:2004mt}) the conversion to a strange star is very likely. 
This conversion process releases a huge amount of energy ($\mathcal{O} \sim 10^{53}\,\mathrm{erg}$) \cite{Bombaci:2000cv}, 
mainly in a powerful neutrino burst \cite{Pagliara:2013tza} that can possibly result in a short gamma-ray burst.  
Thus a way to produce strange stars is through mass accretion onto neutron stars in binary systems \cite{Berezhiani:2002ks,Wiktorowicz:2017swq} or during the spin-down of a rapidly rotating neutron star \cite{Bhattacharyya:2016kte}. 
By this mechanism, ordinary metastable neutron stars could be converted into strange stars and these two families of compact stars could coexist in the universe. 
It is important to emphasize that all the present observational data and our present experimental and theoretical knowledge of the properties of dense matter do not allow us to accept or exclude the validity of the Bodmer--Witten hypothesis and hence the existence of strange stars and the possibility of having two coexisting families of compact stars. 

Matter ejected during the merger of two strange-quark stars is expected to form strangelets \cite{Banerjee:2000ye, Sandweiss:2004bu}, namely strange quark matter droplets with variable baryon number \cite{Berger:1986ps, Bucciantini:2019ivq} due to complex mechanisms as fragmentation or evaporation. Strangelets might also be produced in heavy ion collision experiments \cite{NA52NEWMASS:1996uce, E886:1996ufd} and might be formed from the collisions of energetic cosmic rays with the Earth's atmosphere \cite{Madsen:2002iw}. There has been no unambiguous detection of strangelets, but few candidates events from balloon-borne detectors have been found \cite{Price:1978, Ichimura:1990ce, Saito:1990}. Among the currently running experiments searching for strangelets in cosmic rays are the space-based AMS-02 Spectrometer and the Lunar Soil Strangelet Search (LSSS) collaboration. It is therefore interesting to predict the strangelets flux near the Earth and to compare it with experimental observations.

Previous works computed the production rate in our Galaxy from strange-star mergers: Bauswein \emph{et al.}~\cite{Bauswein:2008gx} performed several binary strange star merger simulations\footnote{Their strange quark matter EOS was derived from the MIT Bag Model, so that the only free parameter is the Bag Constant.} and found a population-averaged ejecta mass of $\sim 10^{-4} \, M_{\odot}$. By assuming a Galactic merger rate of $10^{-5} - 10^{-4} \, \mathrm{yr^{-1}}$ (based on the study of short and long gamma-ray bursts from Belczynski \emph{et al.} \cite{Belczynski:2008uk}), they found a Galactic production rate for strangelets of $\dot{M}_b = 10^{-9} - 10^{-8} \, M_{\odot} \, \mathrm{yr^{-1}}$. A similar computation has been done by Madsen~\cite{Madsen:2004vw} who assumed binary strange star coalescences might emit the same average amount of ejecta than BNS mergers, i.e. $\sim 10^{-5} - 10^{-2} \, M_{\odot}$. The Galactic merger rate was estimated from observations of binary pulsars \cite{Kalogera:2003tn} to be $83.0^{+209.1}_{-66.1} \, \mathrm{Myr^{-1}}$, so that he found a conservative lower limit of $\dot{M}_m = 10^{-10} \, M_{\odot} \, \mathrm{yr^{-1}}$. As discussed in detail below, our simulations suggest that the amount of strangelets produced in mergers could be more than two orders of magnitude larger than estimated in previous works, providing a stronger constraint on the merger rate of strange-star binaries.

From the above discussion, it is clear that the existence of strange stars would have very far-reaching consequence not only for the physics of strong interactions in extreme matter, but also for the many different astrophysical phenomena associated with compact stars. 

We note that although the merging of ordinary neutron stars has been widely discussed in literature \cite{Radice:2020ddv, Bernuzzi:2020txg, Baiotti:2016qnr}, 
the case of binary strange star mergers has been relatively less explored. Only a few works have addressed the general features of the dynamics of this process. They either employed
zero temperature EOSs supplemented with a thermal correction \cite{Zhu:2021xlu,Zhou:2021tgo}, or they included finite temperature effects, but approximated the treatment of gravity using the conformal flatness approximation for the space-time evolution \cite{Bauswein:2009im}. 
In this work we perform the first full general-relativistic hydrodynamic (GRHD) simulations of binary strange star mergers with a microphysical treatment of finite-temperature effects.
We study the difference between this treatment, in which thermal effects are calculated in a way consistent with the zero-temperature equation of state, with those obtained using the zero-temperature EOS supplemented by a phenomenological thermal contribution, like those adopted in previous works \cite{Zhu:2021xlu,Zhou:2021tgo}. 
We also make use of a modified version of the MIT bag model \cite{Weissenborn:2011qu,Bhattacharyyash:2017mdh} that allows for larger maximum masses of strange stars compared to the classical MIT bag model \cite{Farhi:1984qu,Zhu:2021xlu,Zhou:2021tgo}.

The paper is organized as follow. In Sec. \ref{sec:EOS} we describe the modified MIT bag model we employ to obtain the SQM EOS. Sec. \ref{sec_numerical_setup} discusses the numerical setup we adopt and the initial data employed for simulations. An analysis of our results follows in Sec. \ref{sec_results}. In particular in Sec. \ref{sec_merger_dynamics}, we focus on the merger dynamics and in Sec. \ref{sec_GW}, we study GW signatures. Sec. \ref{sec_ejecta} is dedicated to a discussion of dynamical ejecta (\ref{sec_dynamical_ejecta}) from the binary strange star merger and excretion disks (\ref{sec_disk_masses}) for the remnant strange star. The updated estimation of the strangelets flux near the Earth from our binary strange star merger simulations is reported in Sec. \ref{sec_strangelets_flux} along with the comparison with experimental available data. We finally conclude in Sec. \ref{sec_conclusions}. Appendix is dedicated to a couple of important numerical details: the extrapolation of our SQM EOS to very low densities (Appendix \ref{app:extrapolation}) and the display of the tests that we perform on isolated, non-rotating, bare strange star by solving the Tolman–Oppenheimer–Volkoff (TOV) equations (Appendix \ref{app:single_star}).

Throughout this paper we adopt a space-like signature $(-,+,+,+)$ with Einstein's convention for summation over repeated indices. Unless otherwise stated, all quantities are expressed in geometrized units, i.e.,  $G = c = 1$.

\section{The equation of state for strange quark matter}
\label{sec:EOS} 

\begin{figure} 
\includegraphics[width=\columnwidth]{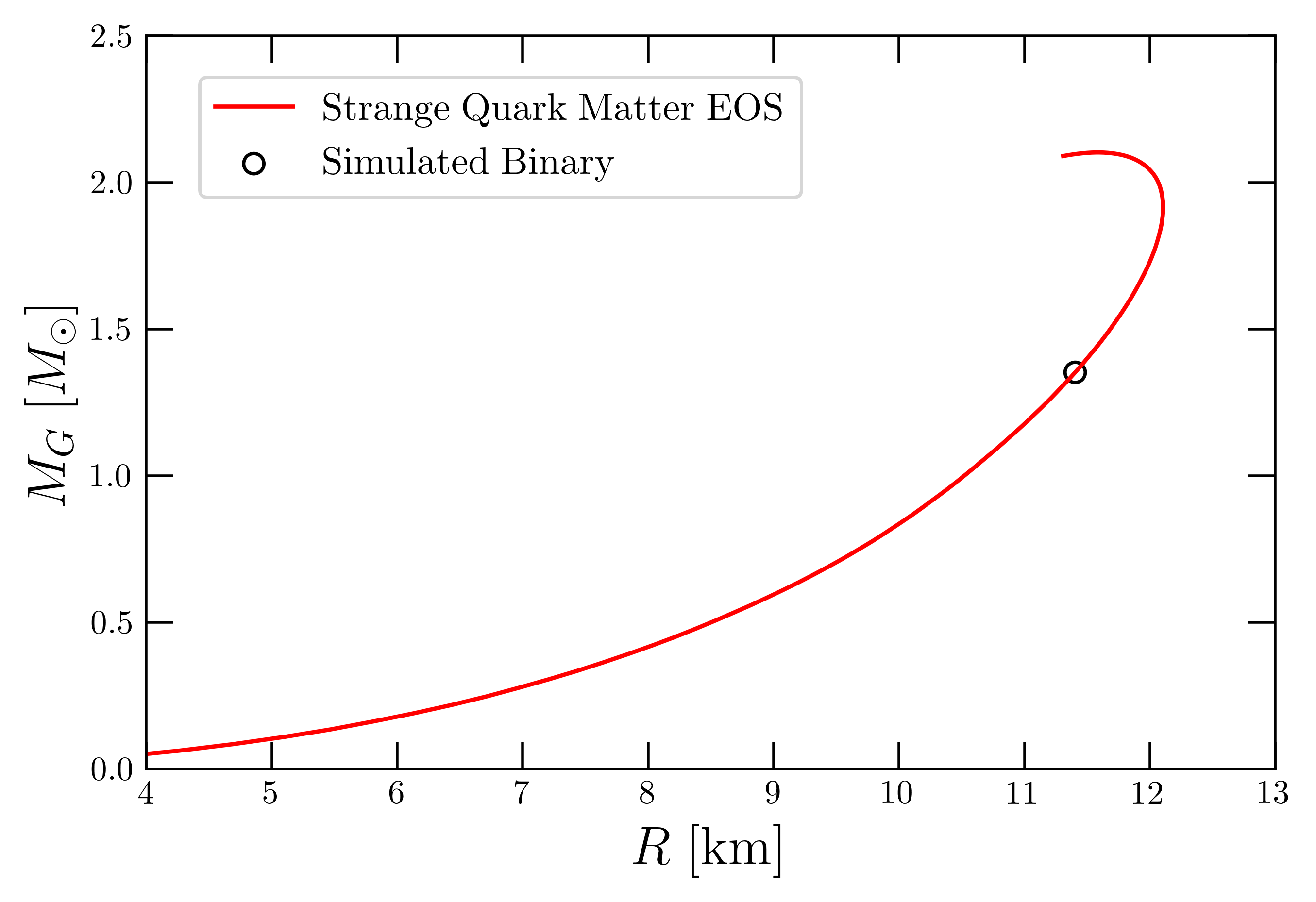} 
\caption{\label{fig:mass-radius} Mass-radius relation for isolated, non-rotating, bare strange stars computed solving the TOV equations for the cold SQM EOS. The black circle denotes the configuration of the equal mass ($m_1 = m_2 = 1.361\, M_\odot$) binary strange star system whose merger is simulated in this work.}
\end{figure} 

The EOS for strange quark matter that includes the effects of gluon-mediated QCD interactions between quarks up to $\mathcal{O}(\alpha_{\rm s}^2)$, can be written \cite{2001PhRvD..63l1702F, Weissenborn:2011qu, Alford:2004pf} in a straightforward and easy-to-use form similar to the popular version of the MIT bag model EOS \cite{Farhi:1984qu}. The grand canonical potential per unit volume takes the form 
(in the units where $\hbar = 1$ and $c = 1$)   
\begin{equation}
\label{eos}
\Omega = \sum_{i=u,d,s,e} \Omega_i^0 +  \frac{3}{4\pi^2}(1 - a_4)\Big(\frac{\mu_{\rm b}}{3}\Big)^4 + B_{\rm eff} \, ,
\end{equation}
where $\Omega_i^0$ is the grand canonical potential per unit volume for {\it u}, {\it d}, {\it s} quarks and electrons modeled as ideal (non-interacting) relativistic Fermi gases individually. 
The second term takes into account the perturbative QCD corrections up to $\mathcal{O}(\alpha_s^2)$ \cite{2001PhRvD..63l1702F, Weissenborn:2011qu, Alford:2004pf}. It represents the degree of deviation from  an ideal relativistic Fermi gas, with the parameter $a_4 = 1$ corresponding to the ideal case. 
The baryon chemical potential $\mu_b$ can be written in terms of the {\it u}, {\it d} and {\it s} quark chemical potentials as $\mu_b = \mu_u + \mu_d + \mu_s$. Finally, $B_{\rm eff}$ is an effective bag constant which phenomenologically describes the non-perturbative aspects of QCD.

\begin{table*}
\caption{A summary of our simulation dataset for two spatial resolutions. $m_1$ and $m_2$ are the gravitational masses of the two strange stars in the binary and $M$ is the total gravitational mass of the binary. $\Lambda$ represents the tidal deformability of the individual star which is the same for both stars in a symmetric binary. $f_2^{\rm peak}$ represents the dominant postmerger peak frequency of the $\ell=2$, $m=2$ mode. $t_{\rm BH} - t_{\rm merg}$ is the time of collapse to a black hole relative to merger.}
\label{Table:summary} 
\newcolumntype{C}[1]{>{\centering\arraybackslash}p{#1}}

\begin{tabular}{ C{3.5cm} C{1.4cm} C{1.4cm} C{1.4cm} C{1.4cm} C{2cm} C{1.7cm} C{1.5cm} C{2.4cm} }
\hline
\hline
 EOS & $m_1 \; [\rm{M_{\odot}}]$ & $m_2 \; [\rm{M_{\odot}}]$ & $ M \; [\rm{M_{\odot}}]$ & $\;\Lambda$ & Resolution & ${f_2^{\rm peak}}$ [kHz] & Collapse & $ t_{\rm BH}-t_{\rm merg}$ [ms] \\
 \hline
 Cold $\rm{+}$ $\Gamma$-Law & 1.36 & 1.36 & 2.72 & 723.34 & LR & 2.836 & Yes & 63.35 \\
 Cold $\rm{+}$ $\Gamma$-Law & 1.36 & 1.36 & 2.72  & 723.34 & SR & 2.576 & No & ... \\
 \hline
 Temperature dependent & 1.36 & 1.36 & 2.72 & 723.34 & LR & 2.668 & Yes & 9.00 \\
 Temperature dependent & 1.36 & 1.36 & 2.72 & 723.34 & SR & 2.682 & Yes & 65.40 \\
 \hline
 \hline
\end{tabular}
\end{table*}

To model the temperature dependence from the ideal gas term provided by fermions and antifermions, we compute the corresponding Fermi integrals for a given temperature $T$ and chemical potential $\mu_i$  (see e.g. \cite{1986bhwd.book.....S}):
\begin{eqnarray}
\Omega^0_{i}(T,\mu_i)&=& - \frac{1}{3} \frac{g_i}{2 \pi^2} \int_0^\infty
k^2 dk\,\,k\, v
\nonumber \\
&\times& \left[f (k,\mu_i) + f (k,-\mu_i) \right] \, ,
\label{Omega_0}
\end{eqnarray}   
where $v = k/E_i$ is the particle velocity, with $E_i(k)=(k^2 + m_i^2)^{1/2}$. 
The functions $f(k,\pm \mu_i)$ are the Fermi distribution functions with chemical
potentials for particles ($+ \mu_i$) and antiparticles ($- \mu_i$) given by:
\begin{eqnarray}
 f(k, \pm \mu_i) = \frac{1}{e^{(E_i(k) \mp \mu_i)/T} +1}.
\end{eqnarray}
The degeneracy factor $g_i = 2$ for electrons and  $g_i = 6$ for each quark flavor. 
Additionally, we neglect the temperature dependence of the last two terms in Eq.(\ref{eos}). 

The total entropy density is given by
\begin{equation}
\label{tot_entropy}
  s = \sum_{i=u,d,s,e} s_i\,
\end{equation}
and can be calculated using the ideal Fermi gas approximation for each fermionic species \cite{Fetter-Walecka:book1971}:
\begin{eqnarray}
s_i (T,\mu_i) &=& - \frac{g_i}{2\pi^2} \int_0^\infty k^2 dk
\left[ f (k,\mu_i) \mathrm{ln} f (k,\mu_i) \right.
\nonumber \\
&+& ( 1 - f(k,\mu_i)) \mathrm{ln}(1-f(k,\mu_i))
\nonumber\\
&+&
f(k,-\mu_i) \mathrm{ln} f(k,-\mu_i)
\nonumber \\
&+& \left. ( 1 - f (k,-\mu_i)) \mathrm{ln}(1-f (k,-\mu_i))
\right] \, .
\label{entropy_f}
\end{eqnarray}

Using standard thermodynamical relations, the energy density can be written as:
\begin{equation}
\label{endens}
e = \Omega + \sum_{i=u,d,s,e}{\mu_i n_i} + Ts \, ,
\end{equation}

where $n_i$ is the number density for each particle species which can be calculated as:
\begin{equation}
\label{numdens}
      n_i = - \bigg(\frac{\partial\Omega}{\partial \mu_i}\bigg)_{T,V}
\end{equation}
and the total baryon number density is:
\begin{equation}
\label{n_B}
       n_B =  \frac{1}{3}(n_u + n_d + n_s)\,.
\end{equation}

Weak reactions of the type:
\begin{eqnarray}
  \label{eq:weak_reactions_1}
  d + u \leftrightarrow u + s\,  \\
  \label{eq:weak_reactions_2}
  u + e^-  \leftrightarrow s + \nu_e
\end{eqnarray}  
occurring as the stellar matter is heated and compressed during the merger of the two stars and the evolution of the post-merger object, will change 
the quark concentrations of matter to minimize the energy per baryon of the system. We initialize the binary in such a way the strange quark matter originally satisfies the $\beta$-equilibrium with respect to the weak interactions, i.e.: 
\begin{equation}
      \mu_s = \mu_d  = \mu_u + \mu_e \, ,
\end{equation}
and electrical charge neutrality. However, the evolution is performed assuming a frozen electron fraction $Y_e$ in each fluid element \cite{Hammond:2022uua}, because we neglect the weak reactions that may occur in the strange stars. Nevertheless, because the equilibrium $Y_e$ is only very weakly dependent on density and temperature for strange matter, we do not expect that this approximation will impact our results in a qualitative way. A more detailed study of the impact of weak reactions in strange-star mergers is left to a future work.

In the present work we take $m_e = 0$, $m_u = m_d = 0$, $m_s = 100~\rm{MeV}$, $B_{\rm eff}^{1/4} = 138\,\mathrm{MeV}$ and $a_4 = 0.8$. 
Using these values for the EOS parameters, SQM satisfies the Bodmer--Witten hypothesis and in addition atomic nuclei are stable with respect to their possible decay into droplets of non-strange (i.e. {\it u}, {\it d}) quark matter \cite{Farhi:1984qu,Bhattacharyya:2016kte}. While describing SQM within the stellar volume, the EOS in Eqs. \eqref{eos}--\eqref{numdens} would provide a negative pressure in the atmosphere surrounding the strange stars. We handle this numerical issue by extrapolating the SQM EOS at low densities via a polytropic equation of state. This introduces an artificial floor that, however, has no impact on the binary dynamics, as the pressure outside the star is many orders of magnitude lower than in its bulk. Appendix \ref{app:extrapolation} is dedicated to this discussion.

In Fig. \ref{fig:mass-radius} we show the mass-radius curve for cold, isolated and non-rotating strange stars described by our EOS model. For the structure parameters characterizing the stellar maximum mass configuration we obtain the following values: gravitational mass $M = 2.10\,M_\odot$, baryonic mass $M_B = 2.71\,M_\odot$, stellar radius $R = 11.57\,\mathrm{km}$, central baryon number density ${n_B}_c = 0.924 \, \mathrm{fm^{-3}}$, central density $\rho_c = 1.7625 \times 10^{15} \,\mathrm{g/cm^3}$, and tidal deformability $\Lambda = 22.46$. For the SQM EOS model employed in this work, general relativistic equilibrium sequences of rapidly spinning bare strange stars have been constructed in \cite{Bhattacharyya:2016kte}. In particular, it was shown (see Table 3, first line and figures 3 and 4 of Ref. \cite{Bhattacharyya:2016kte}) that for the case of maximally spinning bare strange stars (i.e. at the mass shedding limit), the maximum gravitational and baryonic masses correspond to $M = 3.032 \; M_\odot$ and $M_B = 3.924 \; M_\odot$ respectively. 

\section{Numerical setup}
\label{sec_numerical_setup}

\begin{figure*}
\includegraphics[width=0.9\textwidth]{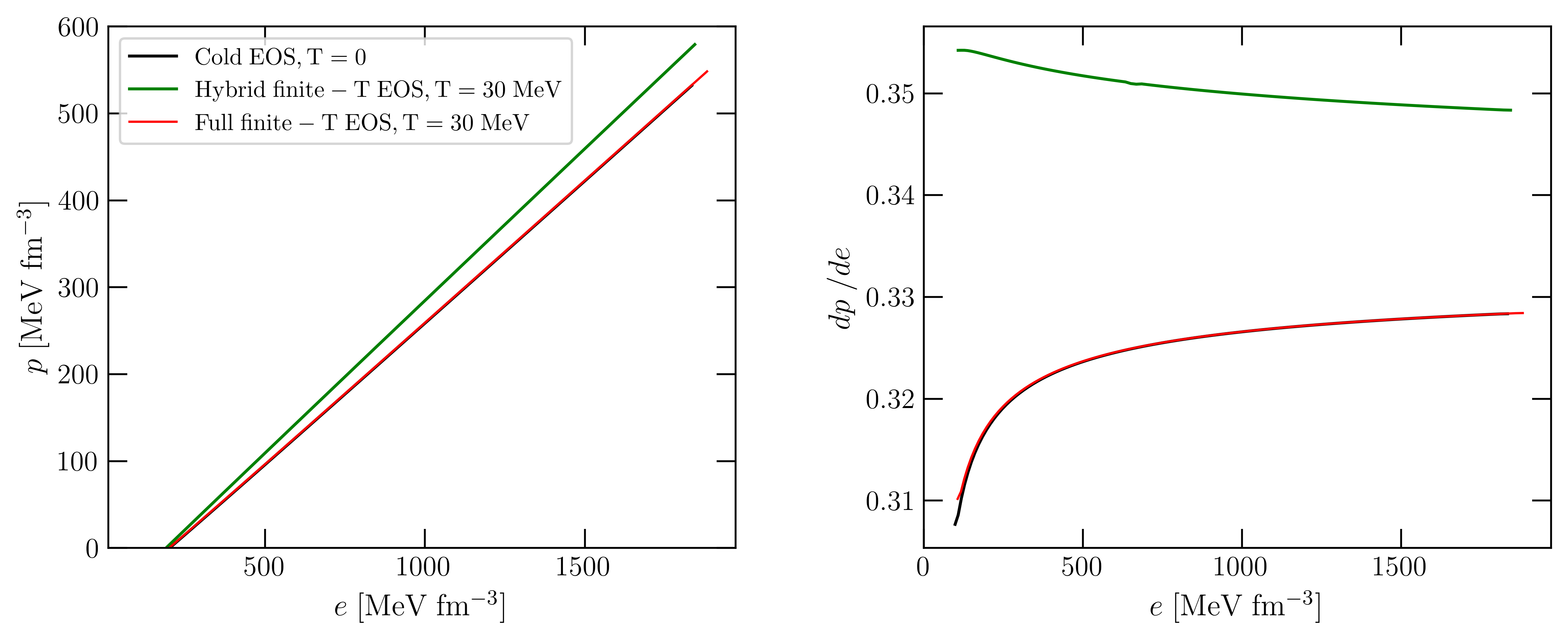}
\caption{\label{fig:EOS_P}
 The variation of pressure with energy density $e$ (left panel) and the pressure gradient $dp/de$ (right panel) for the $\beta-$stable SQM EOS along with the hybrid finite-T EOS and the full finite-T EOS at $30$ \rm{MeV}. The ideal fluid contribution of the $\Gamma$-Law makes the EOS stiffer than the full finite-T EOS at large densities. The larger pressure gradient of the hybrid finite-T EOS suggests enhanced stability for strange stars with this model.} 
\end{figure*}

All the simulations performed in this work are targeted to model a GW170817-like event in the sense that we consider binary strange star configurations with the same chirp mass as GW170817 \cite{TheLIGOScientific:2017qsa}, i.e., $M_{\rm {chirp}} = 1.18 \; M_\odot$. To this aim, we simulate non-rotating, symmetric binaries of total mass $M = 2.72 \; M_\odot$. This choice is aimed at simulating the cleanest scenario before possibly considering the effects of unequal mass ratios.

The initial data are generated using the pseudo-spectral code $\tt{LORENE}$ \cite{Gourgoulhon:2000nn}. We consider irrotational binaries in quasi-circular orbits at an initial separation of $45$ \rm{km}. 
For bare strange stars, we have a huge gradient in density as we move from a tenuous atmosphere to the stellar surface. In fact for the case of the EOS used in this work, the surface density of the star is $\rho_s = 3.58 \times 10^{14} \, \mathrm{g \;cm^{-3}}$ which is $1.33$ times the nuclear saturation density $\rho_\mathrm{nuc}$. Such a steep jump is a consequence of the negative pressure contribution ($-B_\mathrm{eff}$) associated to the bag constant, which makes strange stars self-bound objects with a surface density of the order of the nuclear saturation density. This represents the major numerical challenge to deal with \cite{Zhou:2021upu, Zhou:2021tgo} both in generating initial data and in performing the evolution. For the former, we reduce the relaxation factor for the gravitational potential to $0.05$ in order to control spurious oscillations of density on the stellar surface. We make use of the positivity-preventing limiter of $\tt{WhiskyTHC}$ to handle the surface of the strange stars during their evolution (Appendix \ref{app:single_star}).

The evolution of binary strange star mergers in this work is carried out in full General Relativity using our GRHD infrastructure $\tt{WhiskyTHC}$ \cite{Radice:2012cu, Radice:2013hxh}. The spacetime is evolved using the Z4c formulation \cite{Hilditch:2012fp, Bernuzzi:2009ex} as implemented in the CTGamma \cite{Pollney:2009yz, Reisswig:2013sqa} thorn of Einstein Toolkit \cite {EinsteinToolkit:2023_05}. We employ  the $\tt{Carpet}$ \cite{Schnetter:2003rb, Reisswig:2012nc} infrastructure for adaptive mesh refinement.

We assume the strange star matter to be a relativistic perfect fluid, i.e., a fluid without viscosity, heat conduction or shears. The stress-energy tensor for such a fluid is given by
\begin{equation}
\label{energy-momentum tensor}
    T^{\mu\nu} = (e+p)u^\mu u^\nu + pg^{\mu\nu} \, ,
\end{equation}
where $e$ is the total energy density, $p$ the isotropic pressure, $u^\mu$ the relativistic 4-velocity of the fluid and $g^{\mu\nu}$ the spacetime metric. For numerical reasons, we split the energy density into a ``rest-mass'' and internal energy parts as $e = \rho(1 + \epsilon)$, where $\rho = m_\mathrm{B} n_\mathrm{B}$, $n_\mathrm{B}$ is the baryon number density and $m_\mathrm{B}$ is a mass scale chosen so that $\epsilon \geq 0$. We remark that $m_\mathrm{B}$ does not have a specific physical meaning in the context of strange-quark matter. As described in section \ref{sec:EOS}, we neglect weak reactions in our simulations, i.e., we assume the net lepton number $n_e$ to be conserved. The equations of GRHD are therefore given by the conservation of the baryon number, and the conservation of energy and momentum:
\begin{equation}
\label{conservation_N_B }
    \nabla_\mu (\rho u^\mu) = 0 \, ,
\end{equation}
\begin{equation}
\label{conservation_T_mu_nu}
    \nabla_\mu T^{\mu\nu} = 0 \, .
\end{equation}
The system of equations \eqref{conservation_N_B }--\eqref{conservation_T_mu_nu} is closed by the EOS for strange quark matter which is described in section \ref{sec:EOS}. The flux terms in equations \eqref{conservation_N_B }--\eqref{conservation_T_mu_nu} are reconstructed using a positivity preserving limiter first introduced in \cite{HU2013169} and later implemented in $\tt{WhiskyTHC}$ \cite{Radice:2013xpa}. Remarkably, its usage reduces the numerical oscillations and ensures a very good mass conservation both when simulating single strange stars and for binaries. This evidence along with some details about its functioning are provided in Appendix \ref{app:single_star}.

\begin{figure*}
\centering
\includegraphics[width=0.29\textwidth]{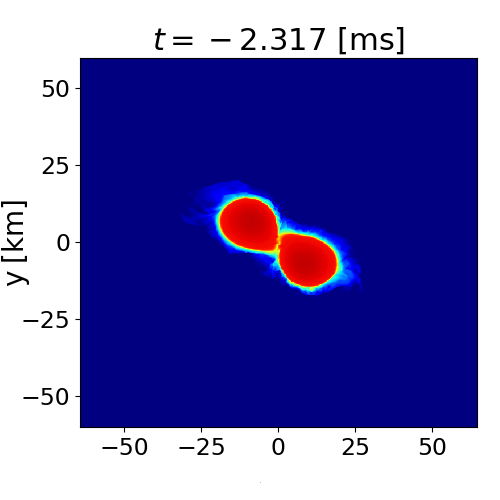}
\includegraphics[width=0.29\textwidth]{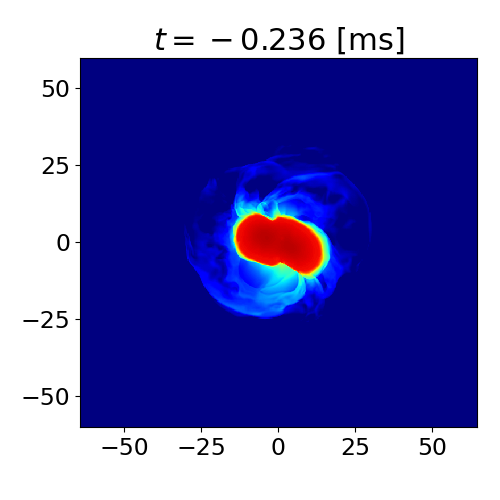}
\includegraphics[width=0.29\textwidth]{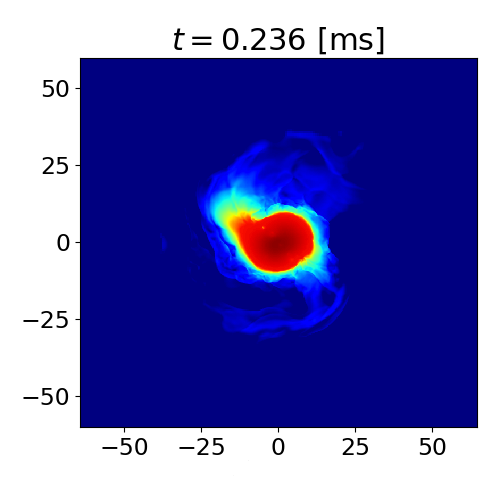}
\includegraphics[width=0.081\textwidth]{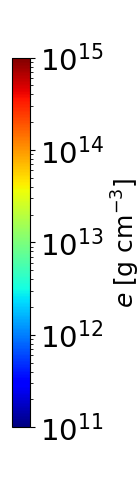}
\includegraphics[width=0.29\textwidth]{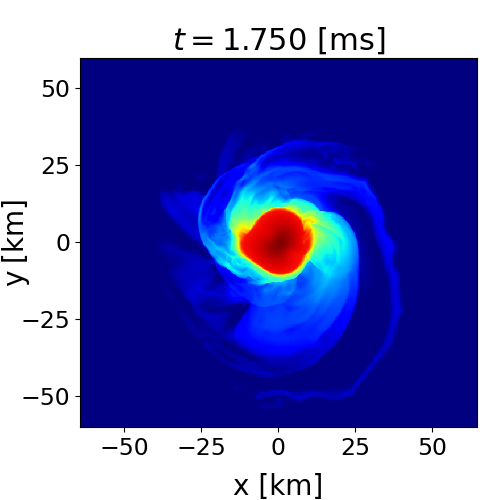}
\includegraphics[width=0.29\textwidth]{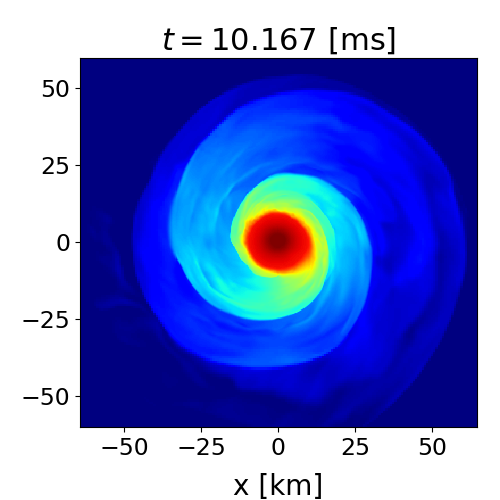}
\includegraphics[width=0.29\textwidth]{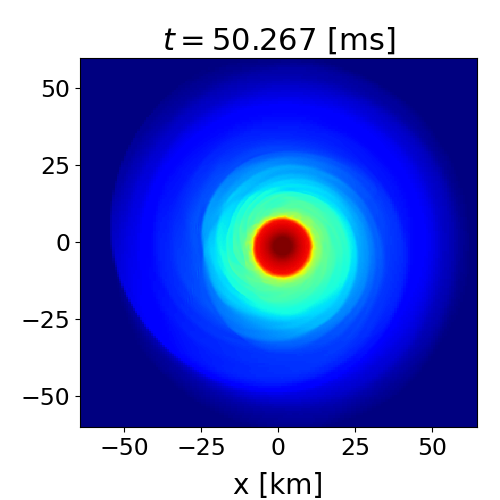}
\includegraphics[width=0.081\textwidth]{Plots/2D/colorbar.png}
\caption{\label{fig:e_fin_SR}2D representative frames of the time evolution of total energy density $e$ across the \emph{xy} plane for a merger of irrotational, bare strange stars evolved with the finite-T EOS (SR). Upper panel: Sketch of the inspiral phase up to the merger. Lower panel: Remnant formation and its dynamics.}
\end{figure*}

All binaries are simulated at two spatial resolutions which we conventionally name as low resolution (LR) and standard resolution (SR). In the finest refinement level, the spatial grid's cell is $\sim$ $180$ \rm{m} long for SR and $\sim$ $250$ \rm{m} for LR. Since no striking differences arise between the two resolutions, we report our results for the most accurate SR unless stated otherwise. Our simulation dataset is summarized in Table \ref{Table:summary}.

An interesting aspect of our work concerns the incorporation of thermal effects into the equation of state for SQM. We compare models of binary strange star mergers with a finite temperature EOS to models where we add an ideal-fluid thermal component to the cold $\beta$-equilibrated EOS. Finite temperature effects are modeled in a consistent manner through the numerical calculation of Fermi integrals for the various thermodynamic variables [Eqs. \eqref{eos}--\eqref{numdens}]. This approach in which thermal contributions are consistently included with the zero-temperature EOS, has only been used in a few simulations of binary strange star mergers \cite{Bauswein:2009im}. In this work, we will refer to numerical simulations that make use of this consistent treatment of thermal effects as the full finite-T simulations. An approximate and still widely used treatment of thermal effects consists of adding the thermal contribution of an ideal fluid to the pressure of the zero temperature EOS:

\begin{equation}
      p = p_{\rm c} + p_{\rm th} \, ,
\end{equation}
where $p_{\rm c}$ is the pressure of the zero-temperature slice of the EOS and $p_{\rm th}$ is the thermal contribution given by the so called $\Gamma-$ law as  
\begin{equation}
      p_{\rm th} = (\Gamma_{\rm th} - 1) \rho\, \epsilon_{\rm th} \, .
\end{equation}
Here $\epsilon_{\rm th}$ is the thermal specific internal energy given by 
\begin{equation}
    \epsilon_{\rm th} = \epsilon - \epsilon_{\rm c}(\rho)
\end{equation}
and the adiabatic index $\Gamma_{\rm th} = 1.7$. We will refer to numerical simulations based on the $\Gamma$-law treatment of the thermal contributions as hybrid finite-T simulations. 

In Fig. \ref{fig:EOS_P}, we report both the variation of pressure and the pressure gradient with total energy density for the zero-temperature EOS as well as for the finite-T and hybrid finite-T EOS at $T = 30 \, \rm{MeV}$. The full finite-T EOS predicts a loss of pressure support at high-densities when compared to the hybrid finite-T EOS thereby making it relatively soft. This manifests as an early collapse of the strange star remnant evolved with the full finite-T EOS as will be discussed in \ref{sec_GW}.

\section{Results}
\label{sec_results}

In this section, we present our results for a merger of binary strange star focusing on the overall dynamics of the remnant, potential signatures on the gravitational wave emission and the dynamical ejection of strange quark matter as a result of the merger.

\subsection{Dynamics of the merger}
\label{sec_merger_dynamics} 

\begin{figure}
\includegraphics[width=\columnwidth]{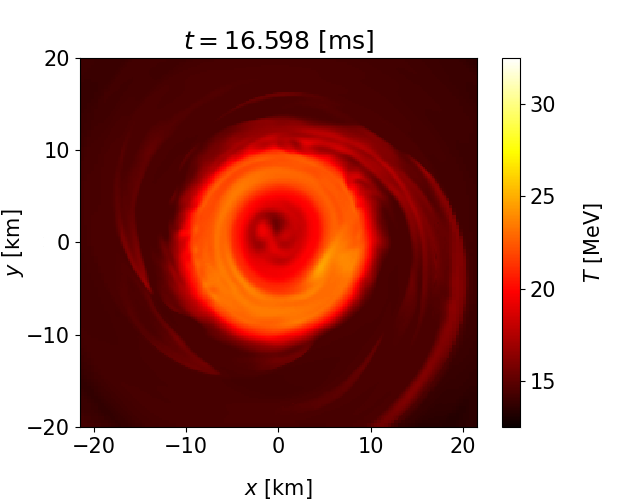}
\caption{\label{fig:T_fin_SR} Instantaneous frame of the temperature inside the remnant formed after the merger of the binary evolved with the full finite temperature EOS (SR).}
\end{figure}

We first present a qualitative overview of the merger dynamics. Since the dynamics are similar for both the thermal treatments, in Fig. \ref{fig:e_fin_SR} we report the evolution of total energy density across the equatorial plane for the full finite-T simulation. We also present the spatial distribution of temperature across the equatorial plane for a representative time in the postmerger (Fig \ref{fig:T_fin_SR}).

The energy density snapshots are reported at instants relative to merger so that the different stages of inspiral, merger and post-merger can be distinguished. Starting from the initial data, the binary emits $\sim$ 7 cycles of GWs in the inspiral, radiating its energy and angular momentum. As the orbit decays, the strange stars become tidally deformed by their mutual gravitational attraction. We define the time of merger $t_{\rm merg}$ as the instance when the $\ell=2, m = 2$ mode of the GW radiation attains a maximum. The merger results in a remnant that is highly deformed and undergoes violent radial pulsations along with differential rotation that source gravitational wave emission in the kilohertz regime (Fig. \ref{fig:strain} in the next section).

\begin{figure}
\includegraphics[width=\columnwidth]{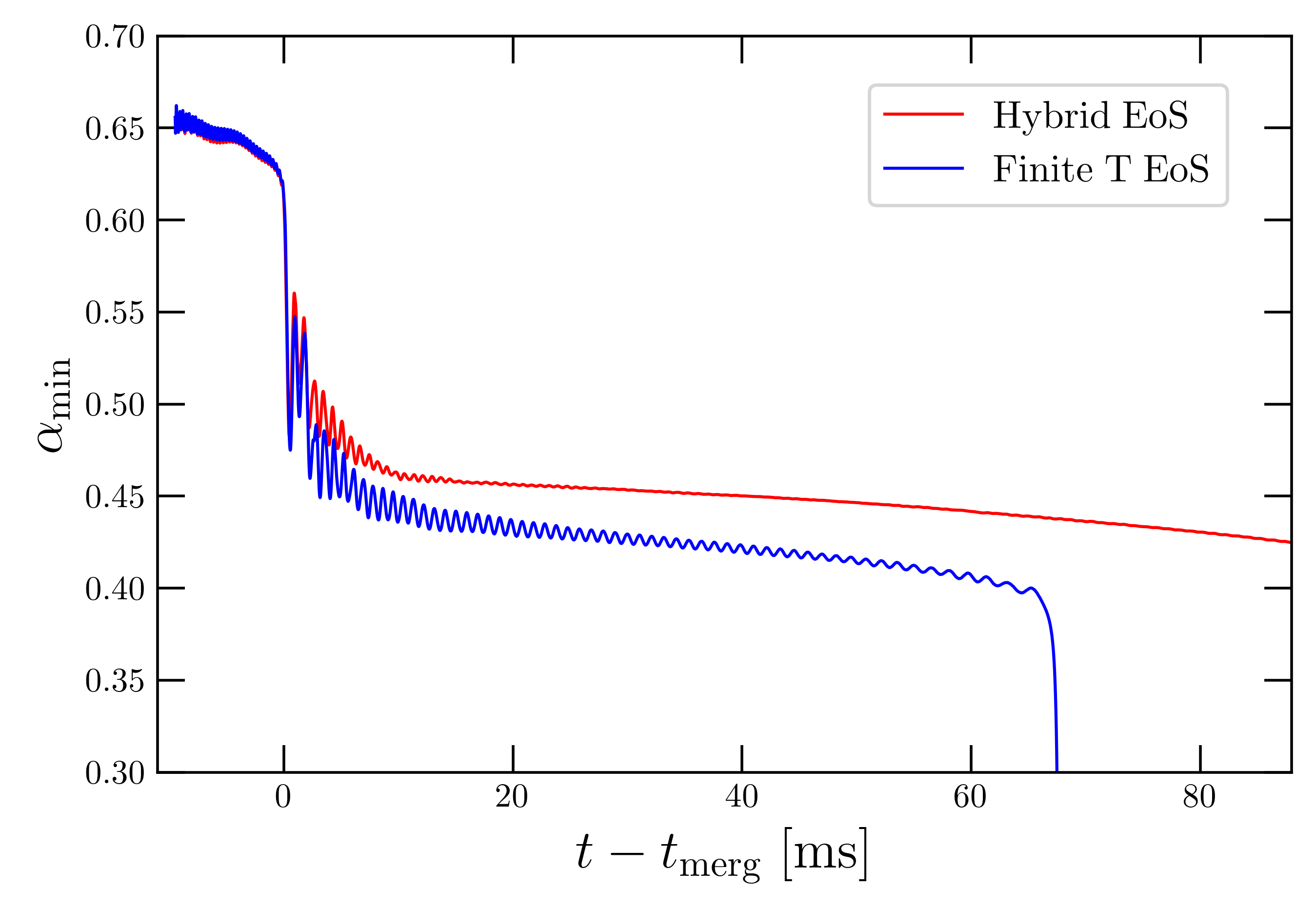}
\caption{\label{fig:min_lapse} Time evolution of the minimum of the lapse function $\alpha_{\rm min}$ with both the hybrid (red) and the temperature-dependent EOS (blue). For the former scenario, we observe the lapse always keeps above $\sim 0.5$, meaning the remnant is a stable body. For the latter, $\alpha_{\rm min}$ drops below 0.3 indicating the formation of an apparent horizon.}
\end{figure}

The remnant thus formed is stabilized by differential rotation over dynamical time-scales $\sim \mathcal{O}(10)$ \rm{ms}, thereby forming a massive and rapidly rotating strange star. We refer the reader to Table \ref{Table:summary} for the eventual fates of the different models we simulate. As we have shown in Fig. \ref{fig:EOS_P}, the full finite-T EOS softens at high densities when compared to the hybrid finite-T EOS. This softening manifests as a difference in the lifetimes of the postmerger remnant. In particular, the binary evolved with the hybrid finite-T EOS at SR is the only one that does not collapse within the simulation time scale. In all other models we find that the remnant is unable to support itself and gives in to gravitational collapse. Consequently, we characterize its lifetime by the quantity $t_{\rm BH} - t_{\rm merg}$ which represents the approximate time from merger when the remnant collapses to a black hole. We identify $t_{\rm BH}$ as the time when the minimum of the lapse function $\alpha_{\rm min} \leq 0.3 $, which for non-spinning binaries is a good approximation. A similar definition for the remnant's lifetime was introduced in Ref. \cite{Bernuzzi:2020txg} and has been utilized in Refs. \cite{Prakash:2021wpz, Kashyap:2021wzs, Bandyopadhyay:2023ohl}. 

We note that the remnant evolved with the full finite-T EOS collapses after 65.40 ms (SR) from the time of merger. This is made explicit in Fig. \ref{fig:min_lapse} where we plot the time evolution of the minimum of the lapse function $\alpha_{\rm min}$. On the other hand, the strange star remnant evolved with the hybrid finite-T EOS remains stable and does not undergo gravitational collapse over the full simulation time of 86.1 ms post-merger. This is evidenced by the near-constant evolution of the minimum lapse. As is expected for remnant lifetimes \citep{Zappa:2022rpd}, we find them to be strongly sensitive to a change in spatial resolution both for the finite-T simulations as well as the hybrid simulations (Table \ref{Table:summary}).

\begin{figure} [H]
\centering
\includegraphics[width=\columnwidth]{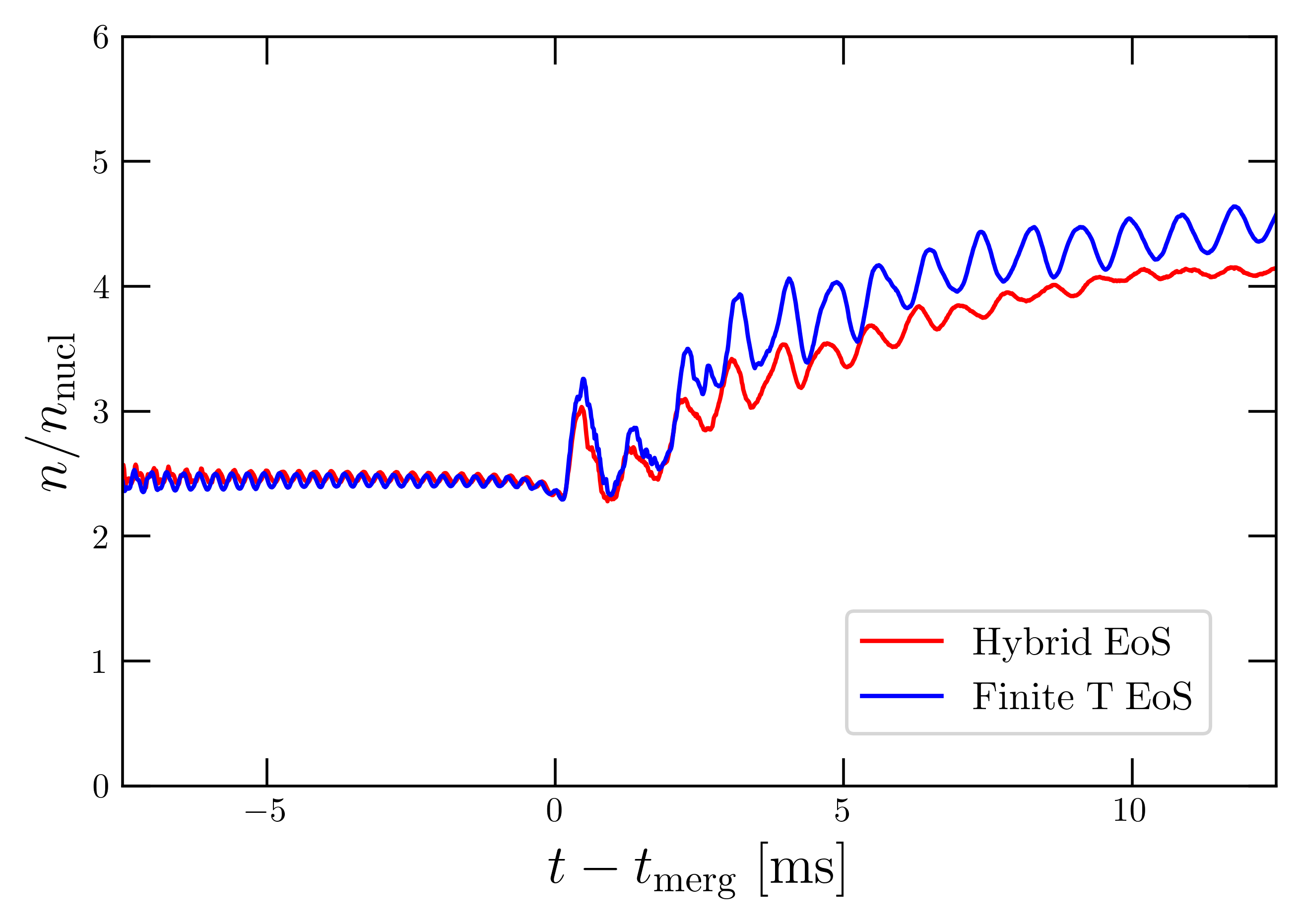} 
\caption{\label{fig:density}Time evolution of the central baryon number density $n_\mathrm{B}$ 
(normalized to nuclear saturation density $n_\mathrm{nucl} = 0.16 \; \rm{fm^{-3}}$) in a merger of strange stars both with the hybrid finite-T (red) and the full finite-T EOS (blue). We observe the full finite-T remnant undergoes more violent oscillations in density owing to its softening at high densities relative to the hybrid EOS. These violent oscillations do not dampen away during the simulation time scale, eventually causing the full finite-T remnant to collapse. On the other hand, oscillations in the hybrid remnant's central density saturate over a time scale of 20 ms, following which the hybrid remnant remains stable.}
\end{figure}

In Fig. \ref{fig:density}, we show the evolution of the central baryon number density $n_\mathrm{B}$ in a merger of strange stars evolved with a hybrid finite-T EOS as well as the full finite-T EOS (SR). During the inspiral, owing to the fact that both EOS treatments have the same cold $\beta-$equilibrated behavior at low densities, the evolution of central density is very similar and undergoes mild oscillations. As the two stars merge, the central density undergoes strong oscillations sourced from the radial pulsations of the postmerger remnant. The difference between the thermal treatment between the two EOSs is reflected in the more violent oscillation of the full finite-T remnant. In particular, these strong pulsations make it unstable towards a gravitational collapse to a black hole. On the contrary, the hybrid finite-T EOS's central density begins to saturate for $t-t_{\rm merg}\gtrsim 10$. As a consequence, the hybrid finite-T remnant can sustain its shape via differential rotation over the time scale of the simulation. The difference in the postmerger dynamics as a result of differences in thermal treatments were also reported in \cite{Bauswein:2009im} where it was found that including non-zero thermal effects in the EOS indeed influences the strange star lifetime.

\subsection{Gravitational waves}
\label{sec_GW}

\begin{figure*}[!ht]
\includegraphics[width=0.8\textwidth]{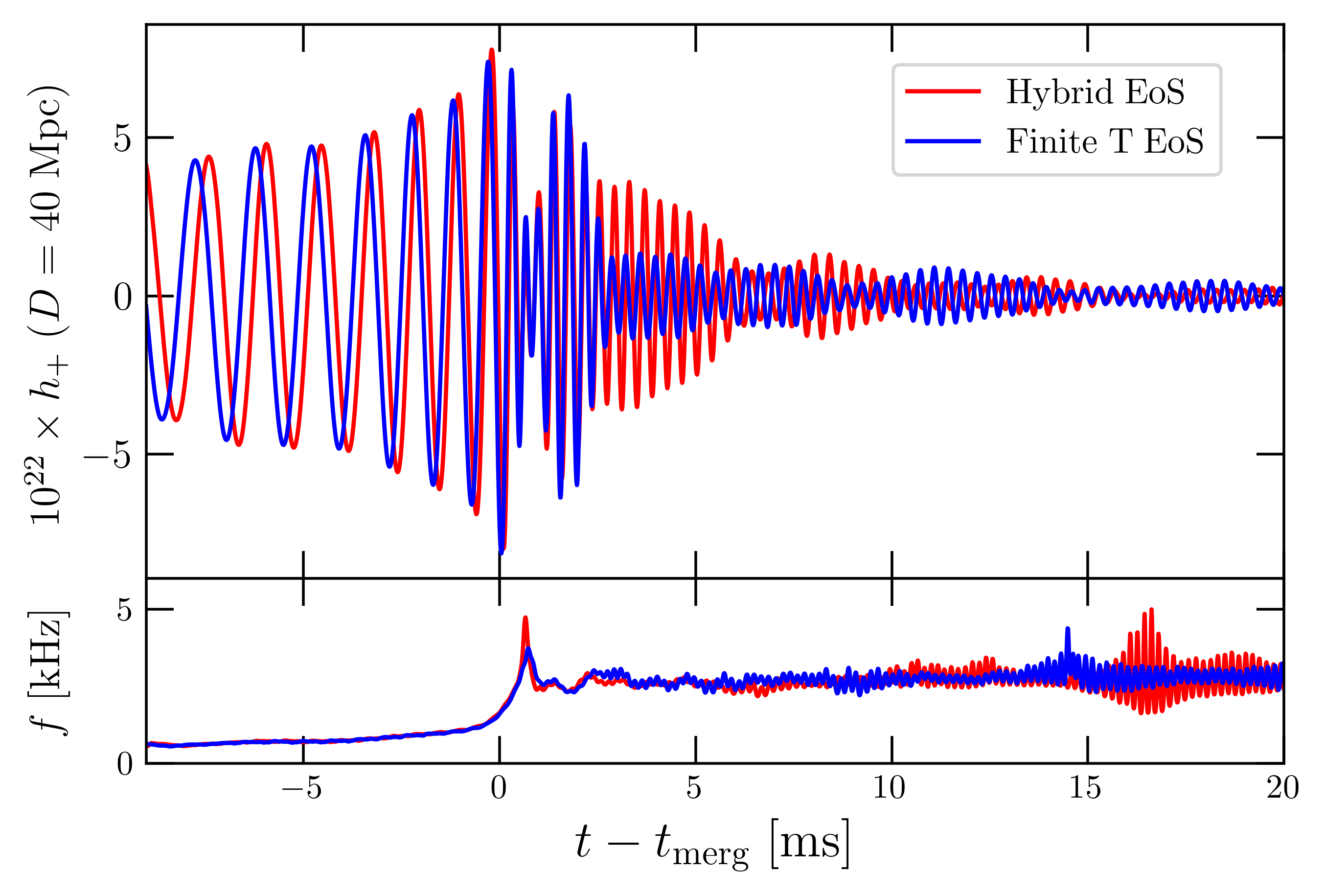}
\caption{\label{fig:strain} Gravitational waves emission from binary strange star mergers with both the hybrid (red) and the temperature-dependent EOS (blue). The strain $h_{+}$ has been oriented so that the merger takes place at $0$ \rm{ms}. \emph{Top panel}: Time evolution of the $\ell=2$, $m=2$ dominant mode of the GW strain. The amplitude of the post-merger signals decreases with time, though with a small modulation. \emph{Bottom panel}: Time evolution of the instantaneous frequency.} 
\end{figure*}

In this subsection, we will present an analysis of the gravitational wave emission from a merger of binary strange stars and compare the evolution from a hybrid EOS finite-T EOS to a full finite-T EOS one. To start with, we employ the Newman-Penrose formalism \cite{Newman:1961qr} to compute the gravitational wave strain $h_{+} -ih_{\times}$ via fast frequency integration \cite{Pollney:2009yz} of the Weyl scalar $\Psi_4$ (measuring the outgoing GWs).

In Fig. \ref{fig:strain} we report the + polarization of the dominant $\ell=2$, $m=2$ mode of the GW strain (top panel) along with the instantaneous frequency (bottom panel) for the three main phases of the coalescence. 

The GW strain is significantly reduced soon after the merger ($\sim2{-}3$~\rm{ms}). Such a reduction is more prominent with the full finite-T EOS. Additionally, we find pronounced modulations in the postmerger amplitude in both the EOS treatments (see \cite{Kastaun:2016elu} for a detailed discussion). These amplitude modulations are anticipated to be a beating pattern caused by the interaction of radial pulsations of the remnant (at a characteristic fundamental mode $f_0 \sim$ $1$ \rm{kHz}) with the rotation of the remnant (at a characteristic postmerger peak frequency $f_2^{\mathrm{peak}} \sim$ $2-4$ \rm{kHz}). Such features in the postmerger waveform morphology have been also been encountered in \cite{Prakash:2021wpz, Prakash:2023afe, Radice:2016rys}.

\begin{figure} [!ht]
\includegraphics[width=0.48\textwidth]{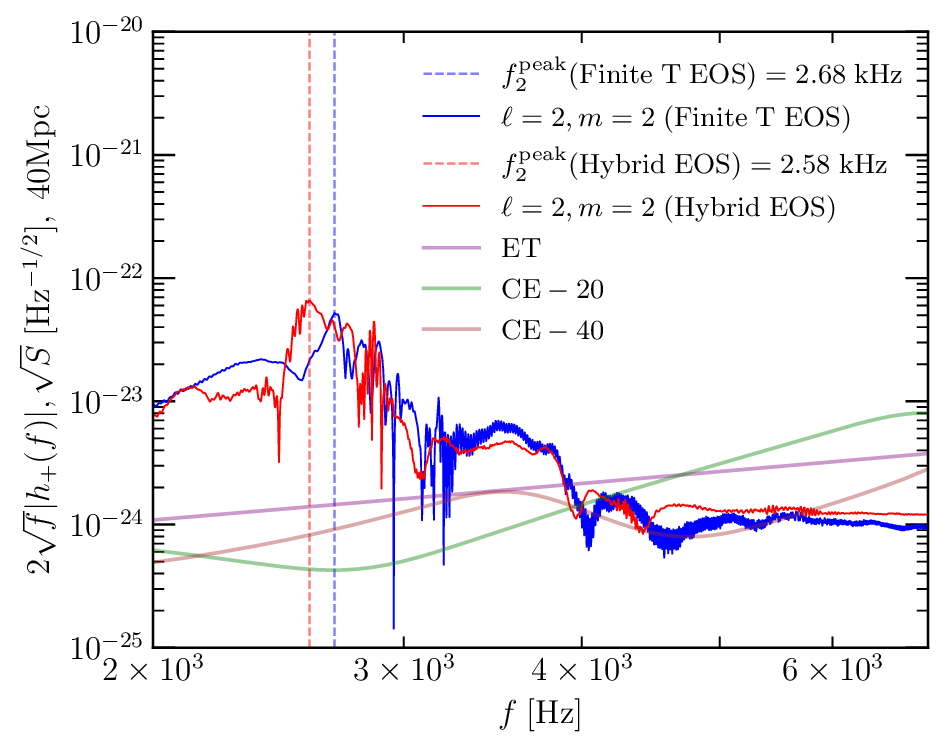}
\caption{\label{fig:psd_together}The amplitude spectral density of the postmerger gravitational wave strain from both the hybrid (red) and the finite temperature EOS (blue). Noticeable are the characteristic postmerger peak frequencies $f_2^{\rm peak}$ between 2 and 4 \rm{kHz}. The finite-T EOS being softer at higher densities naturally predicts a more compact remnant thereby increasing its $f_2^{\rm peak}$. Also reported are the sensitivities of the next generation of GW interferometers with the best detection prospects offered by the $20$ \rm{km} postmerger optimized CE-20 detector.} 
\end{figure}

We report the postmerger amplitude spectral densities of our SR models for the different EOS treatments in Fig. \ref{fig:psd_together}, where we also show the noise curves for the upcoming GW interferometers: the Einstein Telescope (ET) \cite{Branchesi:2023mws, Hild:2010id} and the Cosmic Explorer (CE) \cite{Evans:2023euw, LIGOScientific:2016wof}. The inspiral contribution has been suppressed using a Tukey window to better compare the spectral contributions from the postmerger. We notice that the $f_2^{\rm peak}$ frequencies are different between the two EOS models. This is to be expected because the full finite-T EOS is softer as compared to the hybrid finite-T EOS at the densities and temperatures typically probed during the postmerger. These features in the postmerger spectra provide optimal detection avenues with the next generation of GW detectors. Quantitatively, for a binary strange star merger at a luminosity distance of $40$ \rm{Mpc} (same as that of GW170817), the $20$ \rm{km} postmerger-optimized CE detector will report a postmerger signal-to-noise ratio (SNR) of $38.4$ for the hybrid finite-T model and 30.89 for the full finite-T EOS. This relative decrease in the SNR could perhaps be attributed to the weaker amplitude modulations of the GW strain for the more compact finite-T remnant.

\begin{figure} [!ht]
\includegraphics[width=0.48\textwidth]{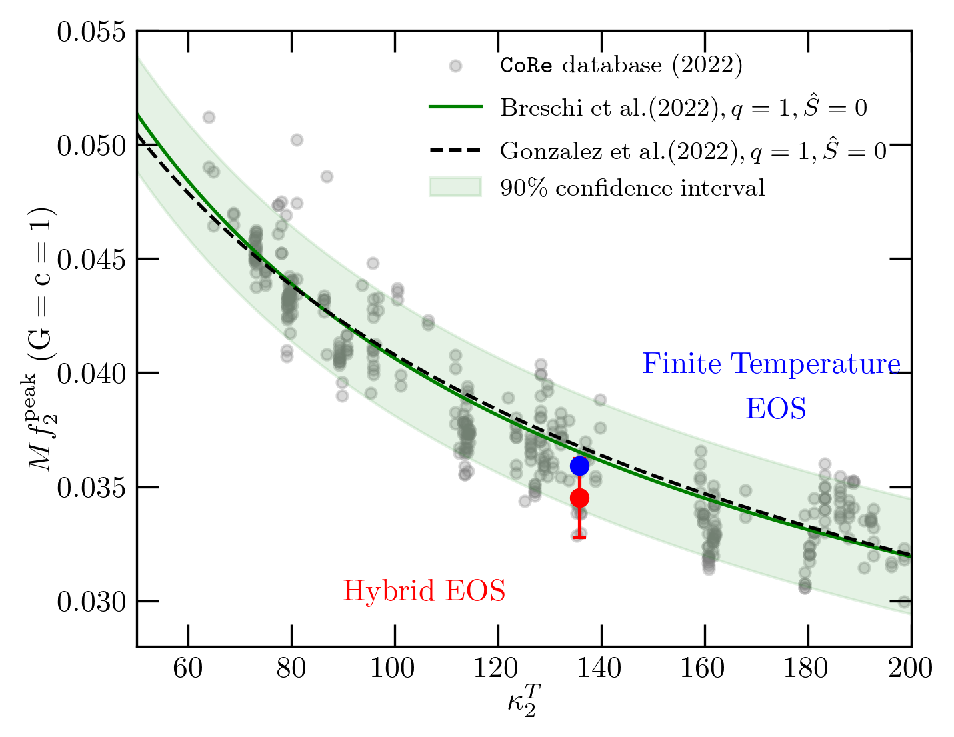}
\caption{\label{fig:qur} Our binary strange star merger models simulated at SR with both the hybrid (red circle) and a finite-T EOS (blue circle) in conjunction with the pre-established quasi-universal relations for BNS mergers provided in Refs. \cite{Breschi:2022xnc, Gonzalez:2022mgo}. The universal relations have been calibrated over 600 Numerical Relativity simulations of neutron star mergers from the $\tt{CoRe}$ database. Up to 90\% credible intervals, binary strange star mergers are degenerate with BNS mergers implying that they are not mutually distinguishable. The error bars provide differences in $f_2^{\rm peak}$ from the corresponding lower resolution (LR) models. Owing to small $\Delta f_2^{\rm peak} \sim \mathcal{O}(10)$ Hz with a change in spatial resolution for the finite-T EOS model, its error bars are not visible on the scale of the plot.} 
\end{figure}

An important question is whether we can differentiate between mergers of neutron stars and mergers of strange stars. There have been a few works in this regard, most notably \cite{Bauswein:2009im} which claimed the differentiability of binary strange star mergers from binary neutron star mergers by citing smaller tidal parameters for strange stars as well as the fact that characteristic inspiral and postmerger frequencies are in general higher for binaries of strange stars. More recently, Ref. \cite{Pradhan:2023pwk} claimed distinguishability between binary strange star and binary neutron star mergers from their respective inspiral signals. The authors made use of the empirical relations between the $f$-mode frequencies (frequency of density perturbations during the inspiral) and tidal deformations and claimed a non-degeneracy between such relations for binary strange star and binary neutron star inspirals. However, such a non-degeneracy could only be established with a weak statistical significance.

Having performed new fully GR simulations of binary strange star mergers, we now report on the differentiability between mergers of strange stars and mergers of neutron stars. In this regard, Zhu \emph{et al.} \cite{Zhu:2021xlu} showed that mergers of strange stars followed the same quasi-universal relations between merger/postmerger frequencies and tidal deformabilities as other hadronic binaries, thereby claiming that it will be difficult to distinguish the two classes of stars.  Our results appear to be in agreement with theirs, as we also find that our strange star merger models follow the pre-established quasi-universal relations between the postmerger peak frequency $f_2^{\rm peak}$ and a tidal parameter $\kappa_2^T$. In particular, in Fig. \ref{fig:qur}, we show this relation for neutron star binaries provided in Refs. \cite{Breschi:2022xnc, Gonzalez:2022mgo} between $f_2^{\rm peak}$ and the tidal parameter $\kappa_2^T$ defined as:
\begin{equation}
    \label{eq:k_2}
    k_2^T = 3 \nu \left [ \left ( \frac{m_1}{M} \right )^3 \Lambda_1 + (1 \leftrightarrow 2) \right] \, ,
\end{equation}
where $\nu = m_1 m_2/M$ is the symmetric mass ratio, $\Lambda_i = (2/3) k_{2,_i} C_i^{-5}$ the quadrupolar tidal deformability of the $i$-th star and, in turn, $k_{2,_i}$ and $C_i^{-5}$ are respectively the Love number and the compactness of the $i$-th star (see also \cite{Breschi:2022xnc}). These relations have been calibrated over $600$ numerical relativity simulations of BNS mergers available publicly via the $\tt{CoRe}$ database \cite{Gonzalez:2022mgo}. We also show (in red and blue circles) our SR models of binary strange star mergers along with error estimates that provide differences in $f_2^{\mathrm{peak}}$ from the corresponding LR simulations. For the binary evolved with the full finite-T EOS, the change in spatial resolution produces a miniscule error of $\Delta f_2^{\rm peak} \sim \mathcal{O}(10)$ \rm{Hz} as compared to the hybrid finite-T model. Both models are within the trends observed for neutron star binaries up to a credible interval of 90\%. Thus in conclusion, up to 90\% credible intervals, it is difficult to distinguish between mergers of strange stars and merger of neutron stars from their postmerger GW emission, at least according to our EOS models.

\subsection{Ejecta}
\label{sec_ejecta}

We shall now investigate the properties of nuggets of strange quark matter, also called strangelets \cite{Bucciantini:2019ivq, Alcock:1986bw} that are ejected in a binary strange star merger. It is noteworthy that the dynamics of strangelets and the consequences of a \emph{contamination} of the Universe are still unclear. In particular, strangelets may stay as SQM if they were absolutely stable or they might evaporate into ordinary nucleonic matter. Also their internal properties, their size and the interactions with the environment may affect their fate. Due to a lack of modeling, we cannot comment upon the possible radioactive decay of ejecta, r-process nucleosynthesis, and the consequent kilonova signatures \cite{Bovard:2017mvn, Vsevolod:2020pak, Metzger:2010sy, Fernandez:2018kax, Hotokezaka:2012ze}. The production of strangelets in SQM mergers was previously investigated by Ref.~\cite{Bauswein:2008gx}. They assumed that mergers of binary strange stars are the only efficient mechanism for the production of strangelets and constrained an average SQM ejecta mass per event to be $\sim 10^{-4} \, M_{\odot}$. Furthermore, they found evidence of a strong correlation between the Bag constant for their EOS model and the ejected mass of SQM implying that a measurement of the cosmic ray flux of strangelets could in principle put observational constraints on the value of the Bag constant.

\subsubsection{Dynamical ejecta}
\label{sec_dynamical_ejecta}

\begin{figure}[!ht]
\includegraphics[width=\columnwidth]{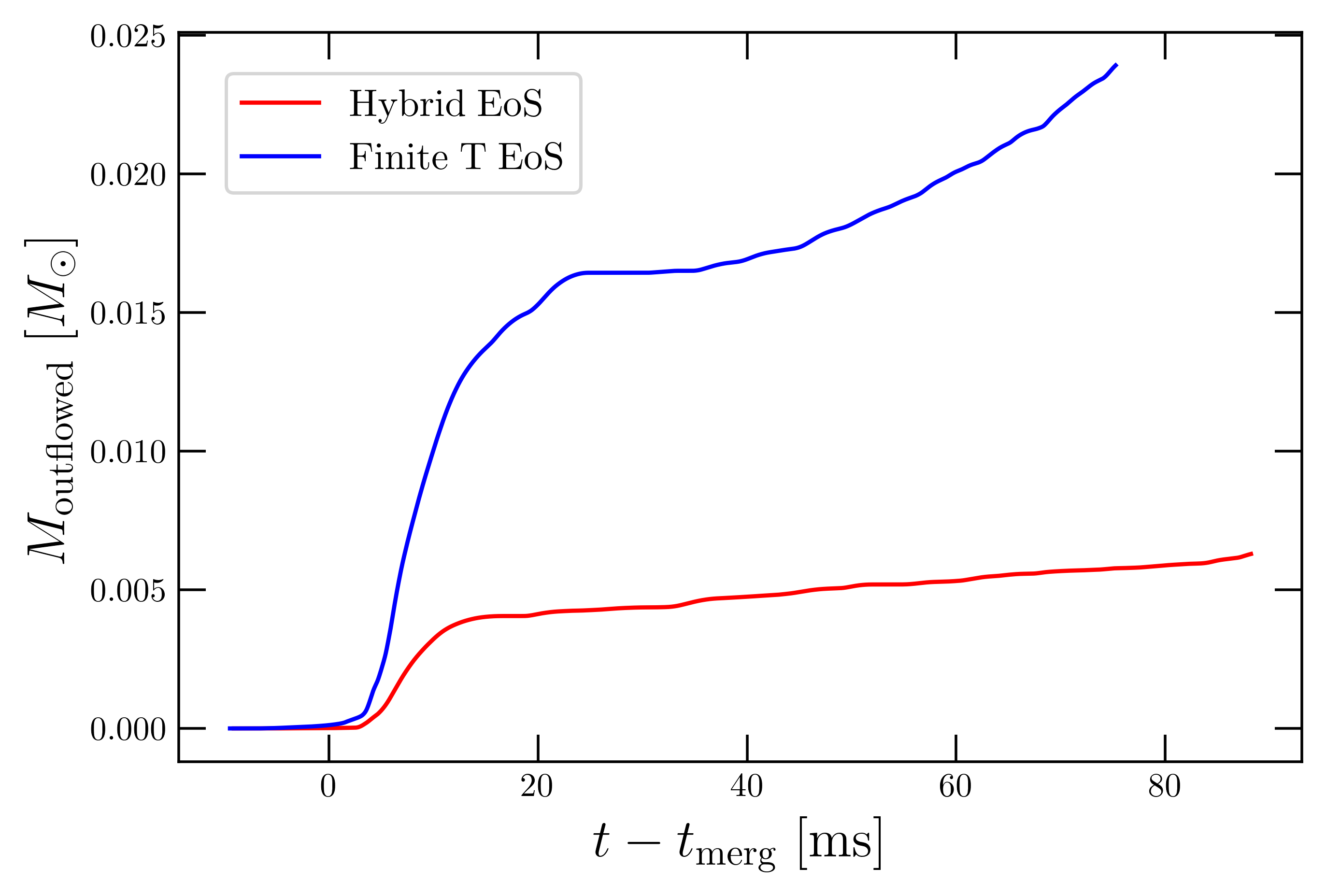}
\caption{\label{fig:total_outflowed_mass}Time evolution of the unbound strange quark matter with both the hybrid finite-T (red) and the full finite-T EOS (blue). There is no significant loss up to few \rm{ms} post-merger, consistent with a symmetric ($q=1$) binary's evolution. The stronger density pulsations of the softer full finite-T remnant in the postmerger are expected to produce more dynamical ejecta.}
\end{figure}

In this subsection, we report our analysis of the strange quark matter ejected from a binary strange star merger on dynamical time scales. To do that, we make use of the geodesic criterion to flag the region of the flow that becomes gravitationally unbound with respect to the remnant. In other words, unbounded SQM follows the condition $u_t < -1$, with $u_t$ being the time component of the 4-velocity of the flow. In Fig. \ref{fig:total_outflowed_mass} we show the mass of SQM that has crossed a fiducial coordinate sphere of $200$ $G \, \mathrm{M_{\odot}} / c^2$ ($\simeq$ $295.34$ \rm{km}). We also refer the reader to \cite{Kastaun:2014fna, Bovard:2017mvn, Nedora:2019jhl} for more discussions on the different criteria for the unbounded matter. 

There is no significant mass loss up to $t_{\rm merg}$ which is usually the case for a symmetric binary's inspiral. This is also in agreement with Bauswein \emph{et al.} \cite{Bauswein:2009im} who found negligible mass transfer in symmetric mergers of both strange and neutron stars. Later on, the lost mass increases monotonically signifying that most of the kinetic energy of the merging stars is indeed converted into thermal pressure causing the ejection of the material.

We find the ejected SQM mass during the postmerger with the full finite-T treatment to be a factor $\gtrsim 4$ more massive than that simulated from the hybrid finite-T EOS. The violent bounces of the full finite-T remnant's core could be responsible for producing more SQM than the hybrid finite-T case as observed in \cite{Radice:2018pdn, Nedora:2021eoj}. That could also be why the amount of dynamical ejecta for the binary evolved with the hybrid finite-T EOS increases in a smooth fashion once the remnant has formed ($\sim$ $10$ \rm{ms} postmerger) whereas the increase in the total outflowed mass is not only larger for the full finite-T models but also increases at a faster rate.

\begin{figure}
\includegraphics[width=\columnwidth]{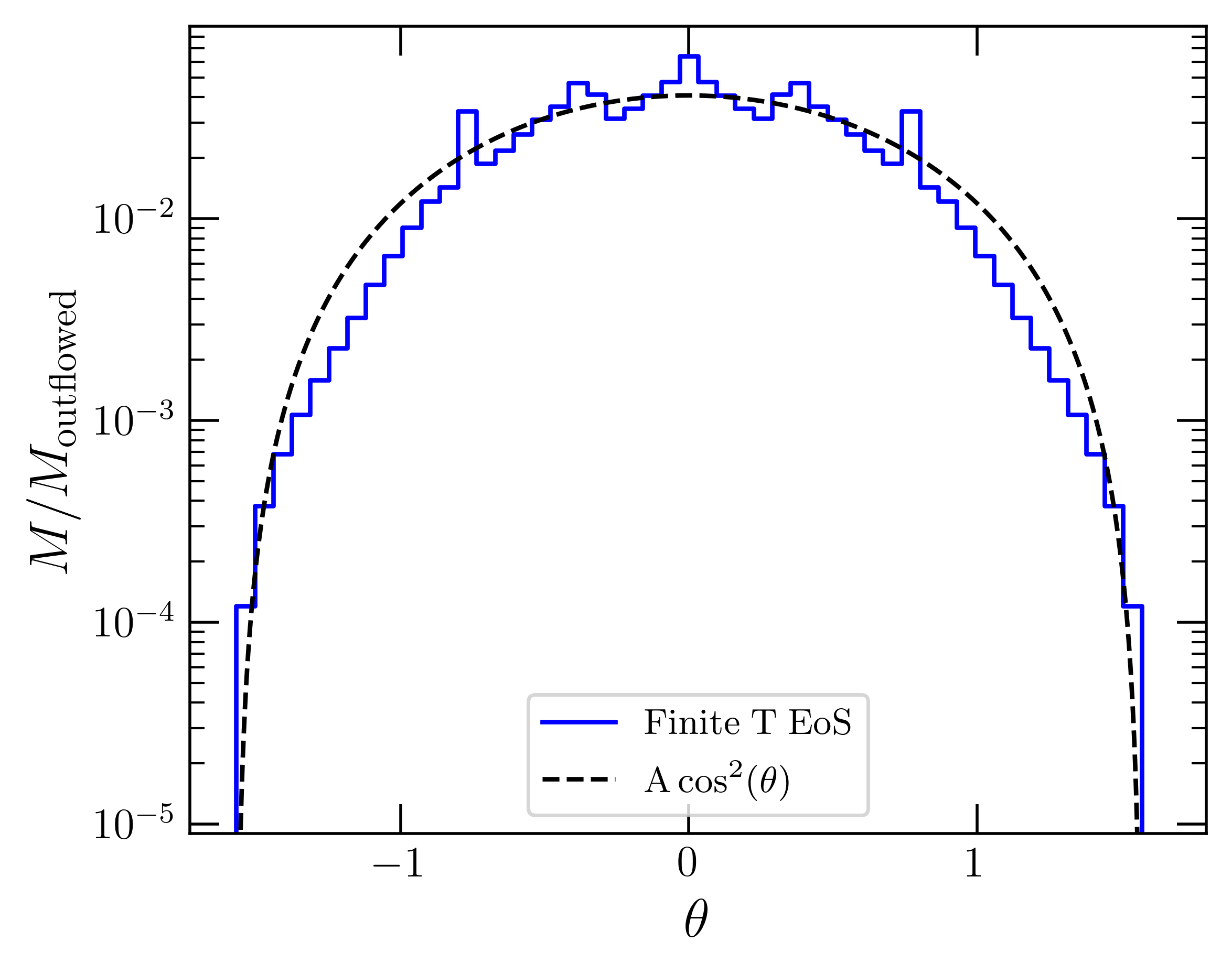}
\caption{\label{fig:hist_theta} Histogram of the ejecta emission angle $\theta$ relative to the equatorial plane for the binary evolved with the full finite-T EOS. The mass represents the amount of ejecta in each bin normalized to the total outflowed mass $M_\mathrm{outflowed}$ (shown in Fig. \ref{fig:total_outflowed_mass}). The analytical approximation $A \, \cos^2(\theta)$ (with $A \sim 0.41$) proposed by Perego \emph{et al.} \cite{Perego:2017wtu} for BNS mergers provides a good fit to the ejecta angular distribution observed in our strange star merger simulations as well.} 
\end{figure}

Quantitatively, we have a dynamical mass loss of $\sim 0.025\ M_{\odot}$ for the binary evolved with the full finite-T EOS; whereas the hybrid finite-T model shows an outflow of only $\sim 0.005\ M_\odot$. Dynamical ejecta from mergers of symmetric binaries \cite{Fujibayashi:2017puw, Radice:2020ddv, Shibata:2019wef} are estimated to typically have a mass $\sim 10^{-3} \, M_{\odot}$, though the influence of the nuclear matter EOS is significant. Hence the amount of ejecta for the binary evolved with the full finite-T EOS is about an order of magnitude larger than the typical mass loss for a comparable BNS merger. This is an interesting outcome because intuitively, one would expect a smaller amount of ejecta owing to the surface tension and gravitational attraction for a strange star. Remarkably, similar to what is observed in binary neutron star mergers \cite{Radice:2018pdn}, we obtain that, while this large emission is concentrated close to the equatorial plane, there is no a strong suppression outside it. The histogram in Fig. \ref{fig:hist_theta} illustrates the distribution of the ejecta emission angle $\theta$ relative to the orbital plane for the binary evolved with the full finite-T EOS. The mass in each bin is normalized to $M_\mathrm{outflowed}$, and we also include the approximation $F (\theta) = A \, \cos^2(\theta)$ proposed by Perego \emph{et al.}~\cite{Perego:2017wtu} to analytically describe the angular distribution of ejecta for BNS coalescences. In our case, the normalization implies $A \simeq 0.41$. The two curves show good agreement, meaning that the strangelets are not preferentially emitted near the equatorial plane and that the approximation in \cite{Perego:2017wtu} even fits well binary strange star mergers data.

Our findings are in tension with those of Bauswein \emph{et al.}~\cite{Bauswein:2009im}, who simulated binary strange star mergers using Smoothed Particle Hydrodynamics. They reported a smaller ejecta mass (0.001 $M_{\odot}$ versus 0.002 $M_{\odot}$ typical of a BNS merger) which is almost entirely confined in the orbital plane. In contrast, our simulations reveal that using a full finite-temperature equation of state results in strong bounces of the remnant’s core, which contribute to a larger and more diffuse emission of dynamical ejecta. The discrepancies between our findings and those of Bauswein \emph{et al.} \cite{Bauswein:2009im} may arise from the different SQM equations of state adopted, as well as the distinct numerical codes employed.

\subsubsection{Accretion Disks}
\label{sec_disk_masses}

We move on to describe the properties of strangelets that are squeezed out of the collisional interface in a binary strange star merger but are still gravitationally bound. Differently from dynamical ejecta, these nuggets of SQM do not possess enough energy to move indefinitely away from the system; instead they form the accretion disk around the remnant. For BNS coalescences, the nature of disks has been widely discussed and is still under investigation. A number of previous works \cite{Shibata:2006nm, Kastaun:2014fna, Hanauske:2016gia} obtained the angular velocity of the gravitationally bound material approximately scales as $r^{-3/2}$, meaning the disk can be assumed to be Keplerian. However, in a recent study \cite{Camilletti:2024otr} a larger set of numerical simulations has been analysed, and it has been suggested that the rotational profile of the disk may rather be characterized by a constant specific angular momentum. Here, we follow the convention in \cite{Radice:2017lry} and we compute the mass of the accretion disk as a function of time from the integral

\begin{equation}
    \label{eq:disk_mass}
    M_{\mathrm{disk}}=\int_V \sqrt{\gamma} \, \rho \, W \, d^3x \,
\end{equation}

over the 3D volume $V$, where $\gamma$ is the determinant of the spatial metric $\gamma_{ij}$, $W$ is the Lorenz factor and the rest mass of strange quark matter should be $\rho < 10^{13} \, \rm{g~ cm^{-3}}$ (larger values of $\rho$ indeed identify the central part of the remnant and no longer the disk). Furthermore the SQM lies outside the apparent horizon (if one has already been formed), i.e., $\alpha \geq 0.3$.

\begin{figure}[!ht]
\includegraphics[width=\columnwidth]{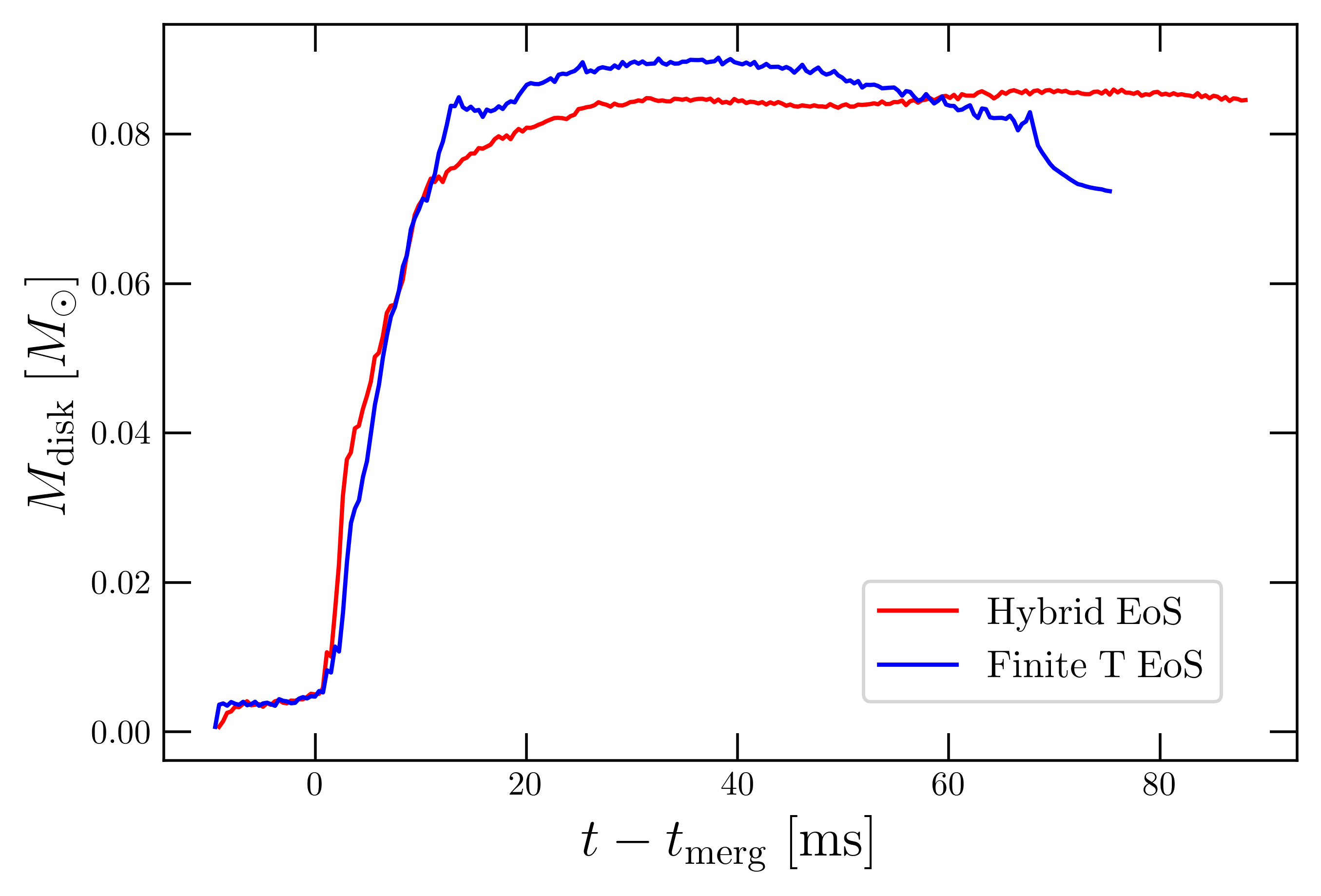}
\caption{\label{fig:disk_mass} Time evolution of the disk mass with hybrid (red) and finite temperature EOS (blue). The stable remnant of hybrid simulation has a more massive disk which remains almost constant for the whole run. Finite-T simulation shows instead a decrease in $M_{\mathrm{disk}}$ once the BH is formed.}
\end{figure}

We report in Fig. \ref{fig:disk_mass} the time evolution of disk mass as described above for both kind of binary strange star merger simulations. First of all, we find that the amount of matter which is gravitationally bound within the disk is greater than dynamical ejecta, though they are of the same order and so the relative difference is not so large (quantitatively it is $\sim 0.07\ M_{\odot}$ for the binary evolved with the hybrid EOS and $\sim 0.05\ M_{\odot}$ with the full finite-T EOS). The mass of the disk usually dominates dynamical ejecta also in neutron star coalescences \cite{Radice:2018pdn, Fujibayashi:2022ftg} (for symmetric binaries $M_{\mathrm{disk}}$ is estimated to be $\gtrsim 10^{-2} M_{\odot}$), but the former is much greater by at least one order of magnitude. Despite such a small relative difference with the strange stars being involved, we find that both EOSs provide values of disk masses that are comparable to those of BNS mergers. Interestingly, our disks are almost as massive as those reported in \cite{Bauswein:2009im}.

\begin{figure}[!ht]
\centering
\includegraphics[width=\columnwidth]{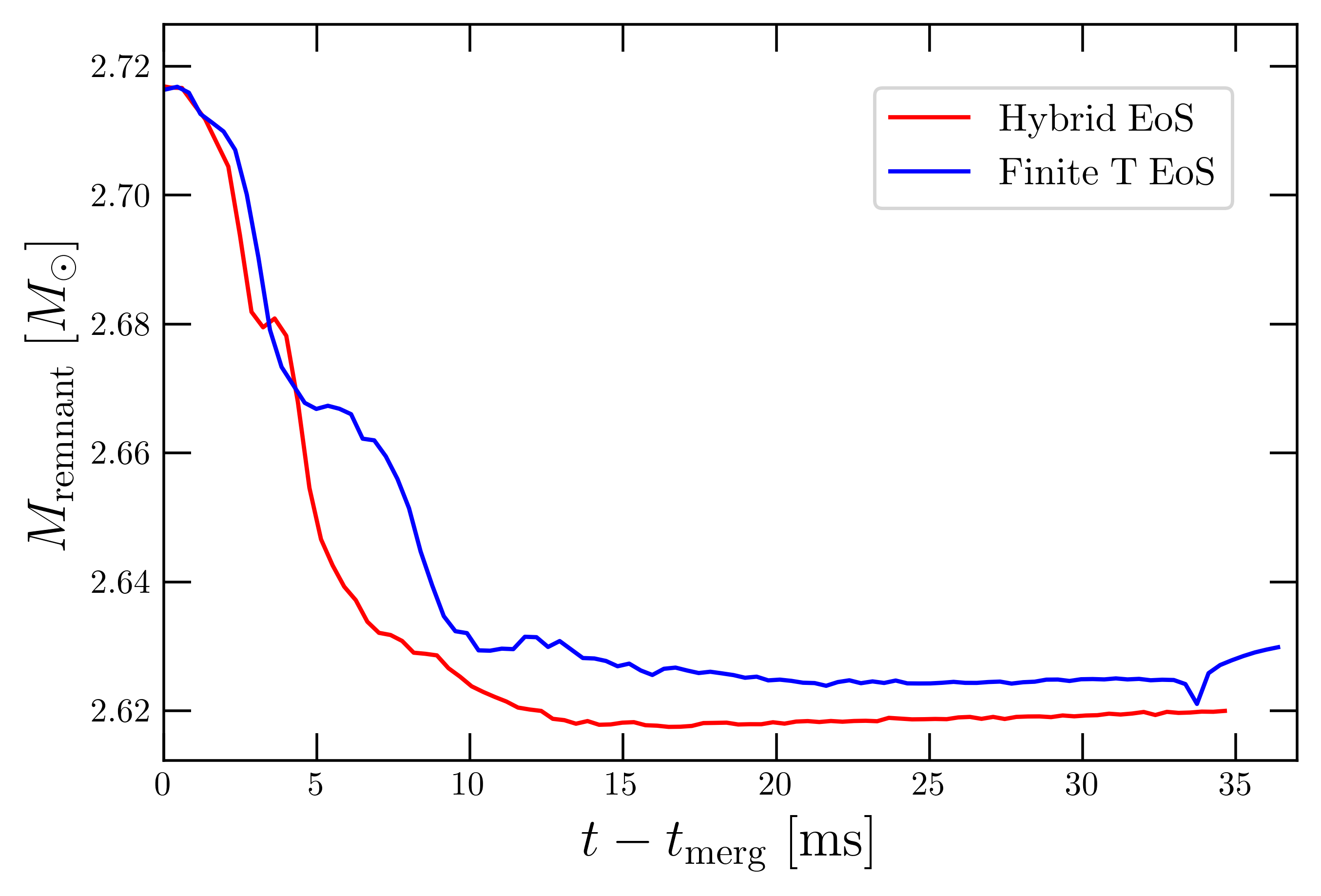}
\caption{\label{fig:remnant_mass} Time evolution of the remnant mass with hybrid (red) and finite temperature EOS (blue).}
\end{figure}

The disk mass of the hybrid model starts to increase after the merger and tends to saturate after about 20 ms. For the finite-T model the disk mass peaks at a value comparable to that of the hybrid simulation, within the intrinsic uncertainty due to finite resolution. Remarkably, we also note a decay in the disk mass for the finite-T model near the time of the formation of an apparent horizon, i.e., $t_{\mathrm{BH}} - t_{\mathrm{merg}} \geq 65.4$~ms. As the collapse begins, some material is accreted back from the disk into the newborn black hole causing a reduction of $M_{\mathrm{disk}}$.
Finally, in Fig. \ref{fig:remnant_mass} we provide an estimation of the mass of the remnant. It is computed as
\begin{equation}
    \label{rem_mass}
    M_{\rm{remnant}} \approx M_{\rm{tot}} - (M_{\rm{outflowed}} + M_{\rm{disk}}),
\end{equation}
where $M_{\rm{tot}} \approx 2.72 M_{\rm{\odot}}$ is the total gravitational mass of the binary. 

At $t_{\mathrm{merg}}$, the remnants formed from the coalescence of both binaries have an initial mass that is only slightly smaller than $M_{\mathrm{tot}}$. Later on, both remnants lose some mass that either feeds the surrounding disks or becomes gravitationally unbound from the system. As for disks in Fig. \ref{fig:disk_mass}, $\sim 15$~ms after merger the masses stabilize, and we find the remnant obtained in the binary evolved with the full finite-T EOS is more massive [in the sense of Eq.~\eqref{rem_mass}]. This is related to the fact that the thermal pressure in the remnant gives in to the gravitational attraction and the remnant collapses to a BH. On the contrary, the ``hybrid'' remnant is not only less massive but also exhibits stronger pressure (because of the stiffer EOS in Fig. \ref{fig:EOS_P}), eventually being able to support the gravitational self-interaction and staying stable for the whole simulation time-scale.

\subsection{Strangelets flux near the Earth}
\label{sec_strangelets_flux}

To estimate the strangelets flux near the Earth, we proceed as in Bauswein \emph{et al.} \cite{Bauswein:2008gx} and Madsen \cite{Madsen:2004vw}. We conservatively assume that each strange-star merger produces $10^{-2} \, M_{\odot}$ of ejecta. Note that in addition to the dynamical ejecta, we can expect part of the disk to become unbound. For simplicity, we work under the assumption SQM is the lowest energy state for matter and all neutron stars are in fact strange stars. However, we remark that scenarios do exist in which strange quark-matter stars and neutron stars can coexist. Recent estimates of the Galactic binary neutron-star merger rates are given by Chruslinska \emph{et al.}~\cite{Chruslinska:2017odi} ($21^{+28}_{-14} \, \mathrm{Myr^{-1}}$) and Pol \emph{et al.}~\cite{Pol:2018shd} ($42^{+30}_{-14} \, \mathrm{Myr^{-1}}$, to within 90\% CL). These rates are consistent with the latest data release from the LIGO-Virgo-KAGRA collaboration \cite{KAGRA:2021duu} that estimates the binary neutron star density merger rate to be $10 - 1700 \, \mathrm{Gpc^{-3}} \, \mathrm{yr^{-1}}$, which would correspond to a local ($z \sim 0$) binary neutron star merger rate density of $48 \,\mathrm{Gpc^{-3} \, \mathrm{yr^{-1}}}$. For our estimate we assume a Galactic merger rate of $\sim 40 \, \mathrm{Myr^{-1}}$, which yields a Galactic strangelets production rate of $\dot{M} = 4 \times 10^{-7} \, M_{\odot} \, \mathrm{yr^{-1}}$. This is at least one order of magnitude greater than Bauswein \emph{et al.} \cite{Bauswein:2008gx}. 

We assume that strangelets undergo the same solar modulation mechanism described in \cite{Madsen:2004vw} when approaching the Earth. By using $\dot{M}$ from our binary strange star simulations, we find their flux is:
\begin{equation}
    \label{eq:flux_strangelets_near_Earth}
    F_{\oplus} \approx 8 \times 10^{8} \,  A^{-1.067} \, Z^{-0.6} \, \Phi^{-0.6}_{500} \, \Lambda' \, \mathrm{m^{-2}} \, \mathrm{yr^{-1}} \, \mathrm{sr^{-1}}
\end{equation}
which is a factor ${\sim} 4000$ larger than Madsen's. In Eq.~\eqref{eq:flux_strangelets_near_Earth} $A$ is the baryon number of strangelets, $Z$ the electrical charge, $\Phi^{-0.6}_{500} = \left( \Phi / 500 \,\mathrm{MeV} \right)$ an electrostatic potential that models the influence of the solar wind when strangelets enter the inner parts of the Solar System. Finally $\Lambda'$ is defined as:
\begin{eqnarray}
    \label{eq:lambda_prime_flux}
    \Lambda' &=& \left( \frac{\beta_{SN}}{0.005} \right)^{1.2} \left( \frac{0.5 \, \mathrm{cm^{-3}}}{n} \right) \nonumber \\
    &\times& \left( \frac{1000 \, \mathrm{kpc^3}}{V} \right) \left( \frac{930 \, \mathrm{MeV}}{m_0 c^2} \right) \, ,
\end{eqnarray}  
where $\beta_{SN}$ is the speed of shocks at which strangelets may be accelerated, $n$ the average hydrogen number density in the interstellar medium, $V$ the effective galactic volume confining cosmic rays and $A m_0$ the rest mass of the strangelet. 

Once emitted from stellar collisions, strangelets experience a complex dynamics. Processes like fragmentation and evaporation into nucleons may significantly alter the internal composition as well as the interactions with the interstellar medium. Furthermore, the substantial lack of accurate models introduces a considerable degree of variability. We therefore provide only a rough estimation of the flux in order to verify the possible implications of our ejecta. In this perspective, we assume $\Phi^{-0.6}_{500} = 1$ and $\Lambda' = 1$ as in \cite{Madsen:2004vw}. Other models that describe the dynamics of the strangelets traveling through the Universe may provide different values, but we do not expect significant variations in the final flux with different assumptions. In contrast, specifying the charge $Z \equiv Z(A)$ and the baryon number $A$ needs more caution: such quantities are even harder to predict and can vary greatly from one model to another. For the former, we again follow Madsen \cite{Madsen:2001bw} who argued that $Z \propto A$ if $A \ll 700$ and $Z \propto A^{1/3}$ if $A \gg 700$ for ordinary strangelets; whereas $Z \propto A^{2/3}$ for color-flavour locked ones. We therefore obtain:
\begin{equation}
    F_{\oplus}^{\mathrm{ord}} \approx 
    \begin{cases}
        \label{eq:flux_ordinary_strangelets}
        2.7 \times 10^9 A^{-1.67} \, \mathrm{m^{-2}} \, \mathrm{yr^{-1}} \, \mathrm{sr^{-1}} & \text{if } A \ll 700 \\
        3.7 \times 10^8 A^{-1.27} \, \mathrm{m^{-2}} \, \mathrm{yr^{-1}} \, \mathrm{sr^{-1}} & \text{if } A \gg 700 \\
        \end{cases}
\end{equation}
and:
\begin{eqnarray}
    \label{eq:flux_CFL_strangelets}
    F_{\oplus}^{\mathrm{CFL}} \approx 2 \times 10^9 A^{-1.47} \, \mathrm{m^{-2}} \, \mathrm{yr^{-1}} \, \mathrm{sr^{-1}} \, .
\end{eqnarray}

To compute the strangelets flux we use Eq.~\eqref{eq:flux_CFL_strangelets}, so we do not need to discriminate between small or large baryon number. Moreover $F_{\oplus}^{\mathrm{ord}}$ and $F_{\oplus}^{\mathrm{CFL}}$ are of the same order of magnitude for all $A$. Hence the quantitative differences are small enough to be negligible for our estimations. As for the baryon number, on one side it is believed that strangelets are emitted with an initial large $A$ (${\sim} \, 10^{43}$ in \cite{Bucciantini:2019ivq}, ${\sim} \, 10^{38}$ in \cite{Madsen:2001bw}) from strange star collisions while on the other their final mass distribution is unclear. Madsen \cite{Madsen:2001bw, Madsen:2004vw} argued that the fragmentation between SQM lumps and the possible evaporation into nucleons can reduce the initial mass by more than 30 orders of magnitude, resulting in \emph{final} strangelets with an average baryon number of ${\sim} \, 10^{3}$. On the other hand, in \cite{Bucciantini:2019ivq} Monte Carlo simulations of scattering and the evaporation mechanism yield much greater \emph{final} SQM fragments. A broad range of fragment sizes is possible, but the majority of strangelets even reach $A \sim 10^{24}$, meaning that the total mass is primarily carried by these larger lumps. Another worth mentioning point is that in Madsen's model [Eq.~\eqref{eq:flux_CFL_strangelets}], the dependence of the flux on $A$ is relatively smooth. An older proposal, not considered here, comes from Wilk \emph{et al.}~\cite{Wilk:2002ez, Rybczynski:2004rr} who predicted $F \propto A^{-7.5}$. Again, further investigations about strangelets emitted from strange star collisions are needed to establish the proper model and that is why we explore different possibilities.

Upper limits on the flux of strangelets near Earth have been reported by a number of experiments. The SLIM detector \cite{Balestra:2006fr, SLIM:2008bwg} has been searching for strangelets in cosmic rays with $A \gtrsim 200 - 600$ depending on the A/Z model. If we select a compatible, small baryon number ($400 \lesssim A \lesssim 1000$), as suggested by Madsen \cite{Madsen:2001bw, Madsen:2004vw}, the corresponding flux turns out to be $3 \times 10^{3} - 3 \times 10^{5} \, \mathrm{m^{-2}} \, \mathrm{yr^{-1}} \, \mathrm{sr^{-1}}$ by using our estimate of the ejecta mass from strange-star mergers. This flux would be larger than SLIM constraints. A larger $A \gtrsim 10^{8}$  would alleviate this tension, though this would no longer be within the range of search. A much greater baryon number ($\gtrsim 10^{20}$), as Bucciantini \emph{et al.} predicted \cite{Bucciantini:2019ivq}, yields a flux of $\lesssim 8 \times 10^{-21} \, \mathrm{m^{-2}} \, \mathrm{yr^{-1}} \, \mathrm{sr^{-1}}$ which is small enough to make strangelets very unlikely to be detected by current experiments. Similarly, the PAMELA collaboration \cite{PAMELA:2015lnr} put the most recent constraints on the flux in the interval\footnote{Most of the strangelets candidates have a small electrical charge.} $1 \leq Z \leq 8$. To stay within these ranges, we may assume $90 \lesssim A \lesssim 250$ (corresponding to $4 \leq Z \leq 8$) that results in a flux of $6 \times 10^{5} - 3 \times 10^{6} \, \mathrm{m^{-2}} \, \mathrm{yr^{-1}} \, \mathrm{sr^{-1}}$, again above the current upper limits. Therefore, experimental constraints on the flux suggest that strange star coalescences may be much rarer than BNS mergers, so that the resulting production rate $\dot{M}$ is significantly smaller than ${\sim} \, 4 \times 10^{-7} $. This means that Bodmer--Witten hypothesis would be disfavored. Alternatively, the SQM assumption could hold under Bucciantini's findings \cite{Bucciantini:2019ivq}: strangelets would conserve a high $A$ and our resulting flux would not lead to any tension. Finally, those mentioned above are not the only experiments that are/have been searching for strangelets. Previous satellite-based \cite{1987ApJ...314..739F, 1989ApJ...346..997B, 1978ApJ...220..719S, 1992Ap&SS.197..121P} and balloon-borne detectors \cite{Price:1978, Ichimura:1990ce, Saito:1990} also constrain the flux to values similar to SLIM and PAMELA whereas some terrestrial searches \cite{Hemmick:1989ns, PhysRevLett.92.022501, PhysRevD.30.1876, PhysRevLett.81.2416} put less strict constraints.

\section{Conclusions}
\label{sec_conclusions}

In this work we performed fully general relativistic simulations of binary strange star mergers by adopting both a cold EOS supplemented by a $\Gamma$-law and a full, finite temperature EOS. In particular, we simulated bare strange stars, composed of strange quark matter up to the surface. We stress our finite-T model is consistent with constraints from QCD. To describe the equilibrium behavior of the degrees of freedom ($u$, $d$, $s$ quarks and electrons) inside a strange star, we have employed an extended version of the MIT Bag model \cite{Weissenborn:2011qu, Bhattacharyyash:2017mdh} which describes the degeneracy pressure of the fermionic species as a non-interacting Fermi gas supplemented by phenomenological descriptions of gluon mediated interactions and non-perturbative aspects of QCD. Our model satisfies the Bodmer--Witten hypothesis \cite{Bodmer:1971we, Witten:1984rs} in that the energy per nucleon is less than that of the most bound $\mathrm{^{56}Fe}$ nuclei. Non-rotating TOV sequences of strange stars constructed from this model are self-bound and can have a very small size $R \sim 4$~km.

We consider two spatial resolutions for each EOS model: Table \ref{Table:summary} summarizes our simulation dataset along with the main physical quantities that we compute to characterize the coalescences.

The dynamics of the merger is qualitatively similar to that found in BNSs. Starting from an initial separation of 50~km (at which we compute the initial data), the strange stars make few orbits around each other before merging. In the postmerger phase, we observe a different fate for the remnant, which collapses in all our simulations but the hybrid model at standard resolution. The remnants formed with the finite-T EOS are relatively more compact and more prone to gravitational collapse to a black hole as compared to remnants evolved with a hybrid EOS. This is attributed to a loss of pressure at high densities in the finite-T EOS when compared to a hybrid EOS.

The postmerger GW signals depict pronounced amplitude modulations as a result of the coupling between the fundamental pulsation mode and the characteristic rotational frequency of the remnant. 
Both models provide one main frequency peak in the power spectral density within the range [2-4] kHz. In particular, the finite-T EOS has as a higher postmerger peak frequency (2.682 kHz), when compared to the hybrid model (2.576 kHz). This is due to the fact that the finite-T EOS is softer. We also find that it will be difficult to distinguish mergers of strange stars from those of NSs using the postmerger peak frequency, since both appear to satisfy the same quasi-universal relations with the inspiral tidal coupling constant.

The study of ejecta reveals some interesting features. We first investigate dynamical ejecta that possess enough energy to become gravitationally unbound from the remnant. They are larger for the finite-T model (reaching $\sim 0.025\ M_{\odot}$ at $\sim 60$~ms postmerger) as a consequence of the intense pulsations in the central density of the remnant. Interestingly, the dynamical mass ejection for the finite-T model exceeds that expected for NS mergers with the same binary masses. On the other hand, the ejecta for the hybrid EOS is smaller ($\sim 0.005\ M_{\odot}$) highlighting the importance of thermal effects and shocks in the mass ejection. In contrast, the disk masses from our models are comparable and, in both cases, saturate at $\sim 0.08\ M_{\odot}$ after $\sim 20$~ms postmerger. Such an amount is similar to the typical one of comparable BNS mergers ($\gtrsim 10^{-2}$ for symmetric binaries). However, the relative difference between gravitationally bound mass and dynamical ejecta is smaller in binary strange star mergers. We indeed find that, in our simulations, disk masses and dynamical ejecta are of the same order, whereas the former is larger than at least one order of magnitude in a comparable BNS merger. We also provide a rough estimation of $M_{\mathrm{remnant}}$, finding the remnant is more massive with the finite-T EOS. The combination of a more massive remnant and a softer EOS providing less thermal pressure causes the eventual collapse to black hole for that model. Interestingly, our findings align with those of Bauswein \emph{et al.} \cite{Bauswein:2009im} regarding the negligible mass transfer during the merger of equal mass strange stars and the mass of the disks that surround the remnants. However, the binary evolved with the temperature-dependent EOS shows a larger outflow ($0.025 \, M_\odot$ versus $0.001 \, M_\odot$ in \cite{Bauswein:2009im}) that is not confined in the orbital plane. These discrepancies might arise due to different numerical codes and equations of state employed.

We obtain revised estimates for the flux of strangelets near Earth from strange-star mergers. We used the approach of Bauswein \emph{et al.} \cite{Bauswein:2008gx} and Madsen \cite{Madsen:2004vw}, but with updated merger rates and using our simulation results. We find a galactic production rate of strangelets to be  $\dot{M} \sim 4 \times 10^{-7} \, M_{\odot} \, \mathrm{yr^{-1}}$, an order of magnitude larger than the estimate of Bauswein \emph{et al.}~\cite{Bauswein:2008gx} and three orders of magnitude larger than the estimate from Madsen~\cite{Madsen:2004vw}.

An accurate estimate of the corresponding flux of strangelets at Earth needs a comprehensive understanding of the dynamics and the interactions of strangelets that travel through the Universe. In particular, the baryon number $A$ may range from $10^2$ to $10^{24}$ depending on the model, significantly affecting the flux.
We discuss two possible scenarios motivated by Madsen \cite{Madsen:2001bw, Madsen:2004vw} and Bucciantini \emph{et al.} \cite{Bucciantini:2019ivq}, who predicted a much different baryon number of strangelets close to Earth ($A\sim 10^{3}$ and $A \gtrsim 10^{20}$, respectively). Experimental upper limits \cite{SLIM:2008bwg, PAMELA:2015lnr} on the strangelets flux suggest that either strange star mergers are rather rare (resulting in a smaller $\dot{M}$, and disfavouring the SQM hypothesis) or strangelets do not significantly fragment.

Finally, we mention some limitations of our work that we leave for further investigations. It would be interesting to simulate a larger number of symmetric and asymmetric strange star binaries to verify how the total mass and/or the mass ratio may affect the dynamics of the merger. Even though it is a reasonable first approximation to neglect weak interactions, the inclusion of such processes would provide a more accurate description of the thermodynamics of strange stars during their mergers. This is expected to be important, given the prominent role of shocks in the ejection of strangelets in our simulations. Finally, our simulations neglected neutrinos and we did not model the thermodynamic properties of the diluted gas of strangelets that might constitute the ejecta from such systems. Improving upon these aspects of our simulations as well as strengthening the estimates of the flux of strangelets near Earth will be the objects of our future work.

\section*{Acknowledgments}
DR acknowledges funding from the U.S. Department of Energy, Office of Science, Division of Nuclear Physics under Award Number(s) DE-SC0021177 and DE-SC0024388, and from the National Science Foundation under Grants No. PHY-2011725, PHY-2020275, PHY-2116686, and AST-2108467.

This research used resources of the National Energy Research Scientific Computing Center, a DOE Office of Science User Facility supported by the Office of Science of the U.S.~Department of Energy under Contract No.~DE-AC02-05CH11231. Computations for this research were also performed on the Pennsylvania State University’s Institute for Computational and Data Sciences’ Roar supercomputer.

\appendix
\section{Details on the numerical schemes}
\label{app:appendix}

In this Appendix, we provide additional insights on the numerical schemes and methodologies used in our study. Specifically, we discuss the extrapolation of the SQM equation of state at low densities and present a series of tests performed on single, non-rotating strange stars to validate our implementation and results.

\subsection{Extrapolation of the SQM EOS at low densities}
\label{app:extrapolation}

\begin{figure}
\includegraphics[width=\columnwidth]{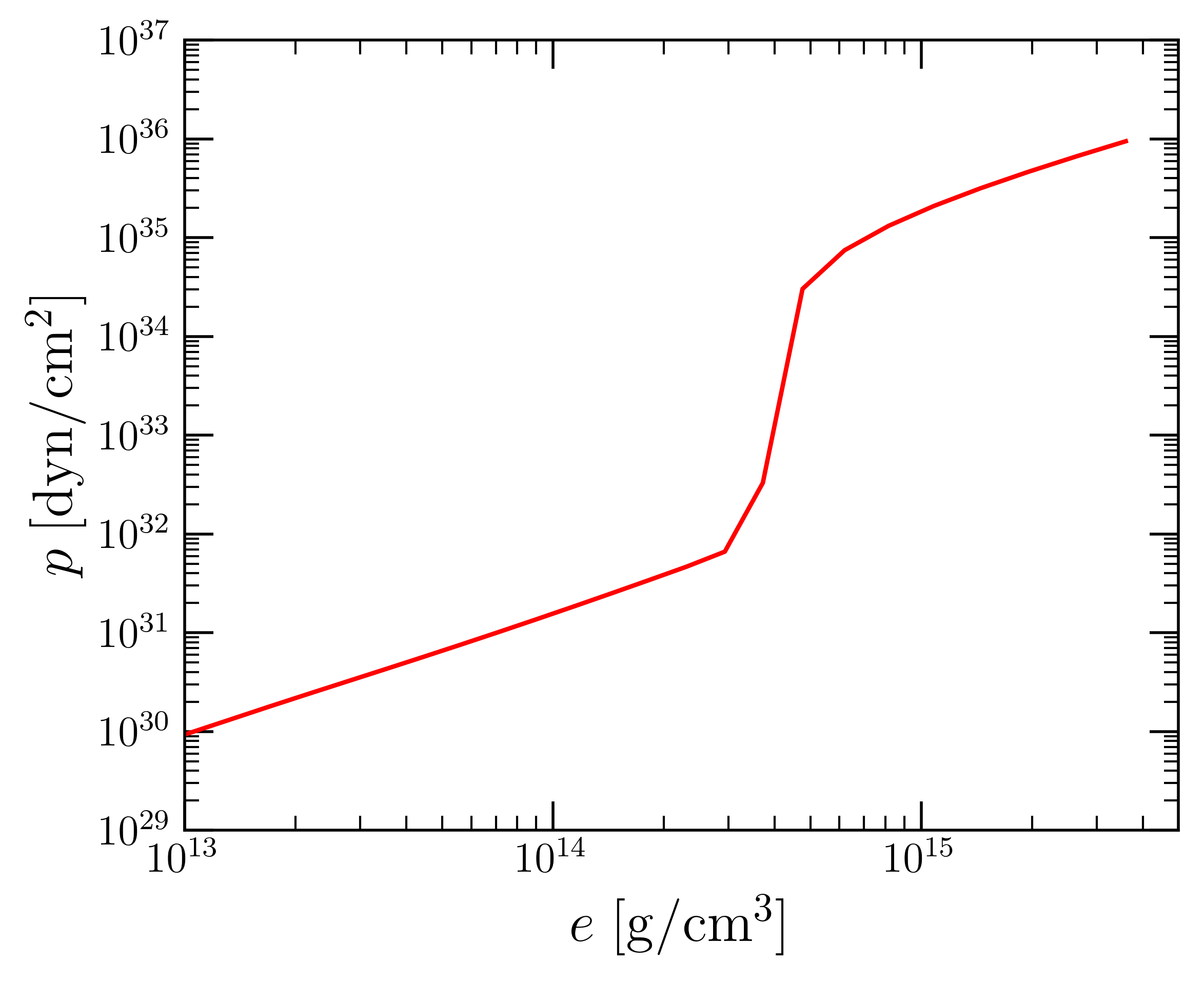}
\caption{\label{fig:P_rho_eos_thermal_new} Pressure vs energy density diagram for the $\beta$-stable SQM equation of state. We use \emph{cgs} units and logarithmic scales to emphasize the sharp discontinuity at the surface of bare strange stars, occurring at $e_{\mathrm{surf}} \sim 4 \times 10^{14} \, \mathrm{g/cm^3}$. Beyond this point ($e < e_{\mathrm{surf}}$), the extrapolation via a polytropic EOS ensures a small but finite pressure even at very low densities.}
\end{figure}

The equation of state formalized in Sec. \ref{sec:EOS} describing the $\beta$-equilibrated SQM is strictly valid at the energy densities achieved inside the star. Here, the free energy contributions of the quarks $u$, $d$, $s$ exceed the negative pressure associated with the bag constant, resulting in a net positive pressure\footnote{Since a strange star is modeled as a perfect fluid, its pressure per unit volume can be straightforwardly derived from Eq. \eqref{eos} as $p = -\Omega$.}. This ensures the stability of the strange star against gravitational collapse.

At the stellar surface the pressure vanishes by definition, i.e., $p(e_{\mathrm{surf}}) = 0$, marking the boundary where quark matter ceases to exist in equilibrium. Beyond this point ($e < e_{\mathrm{surf}}$), the MIT bag model predicts a negative pressure\footnote{The thermal pressure could still provide a positive total pressure in some cases, but this is not guaranteed across all conditions.}, as the confining $B_{\mathrm{eff}}$ dominates over the quark degeneracy pressure at low densities. This implies an unphysical regime where the system cannot sustain itself. Theoretically, the homogeneous quark phase is expected to break down, converting into a gas of zero-pressure strangelets interspersed with vacuum.

Despite the lack of SQM beyond the surface, our numerical simulations require handling these low-density regions due to the presence of the atmosphere. The transition to vacuum poses numerical challenges, as the sharp fluid-vacuum interface can destabilize the computational grid. To address this issue, it is standard practice (also adopted in neutron star simulations \cite{Radice:2013xpa}) to introduce a thin, low-density atmosphere around the star. This artificial layer ensures numerical stability by maintaining a well-defined grid near the surface.

The atmosphere corresponds to a region where the density falls below $e_{\mathrm{surf}}$. In this regime, the SQM EOS cannot be applied, as it would provide negative pressure. At the same time, the physical scenario of a gas of zero-pressure strangelets and vacuum cannot be modeled in our code.

To overcome this limitation, we introduce an artificial small pressure by extrapolating the SQM EOS into the low-density regime using a polytropic equation of state. This approach guarantees that the pressure remains positive.

Fig. \ref{fig:P_rho_eos_thermal_new} illustrates the pressure as a function of total energy density for our EOS. We employ the logarithmic scale to highlight the sharp discontinuity that characterizes the surface of bare strange stars and occurs at $e_{\mathrm{surf}} \sim 4 \times 10^{14} \, \mathrm{g/cm^3}$. Above this density, the SQM populating the star is described by Eqs. \eqref{eos}--\eqref{numdens} while below this density, the extrapolation via a polytropic EOS provides a small but finite pressure that extends into the low-density atmosphere. This enables the accurate simulation of the stellar environment, while ensuring a smooth fluid-vacuum transition and a sound treatment of the floor.

\subsection{TOV tests on isolated strange stars}
\label{app:single_star}

Before simulating the evolution of binary systems, we performed several GRHD tests on bare, isolated, non-spinning strange stars. We report our main outcomes that served us to examine their stability and the proper mass conservation with our SQM EOS. The additional contribution of the strong force in binding strange stars causes a sharp discontinuity in density at the surface that separates the interior of the strange star from the atmosphere. Indeed, while in an ordinary neutron star the rest-mass density decreases with pressure and gradually vanishes moving toward the external layers, in a bare strange star $\rho$ is still very large ($\sim 2-3 \; \rho_{\mathrm{nucl}}$) when approaching $p=0$, i.e. the surface. Handling the sharp discontinuity is a numerical challenge because it may lead to strong shocks and, in turn, to artificial dissipations. The presence of a thin hadronic crust (\cite{Zhou:2021upu}) would smooth the density profile as for neutron stars. However, if a crust is not considered, then it should be shown that no artificial oscillations occur. 
Examining the stability of a TOV configuration by monitoring the evolution of central density serves to establish its stability. 

\begin{figure} [!ht]
\includegraphics[width=\columnwidth]{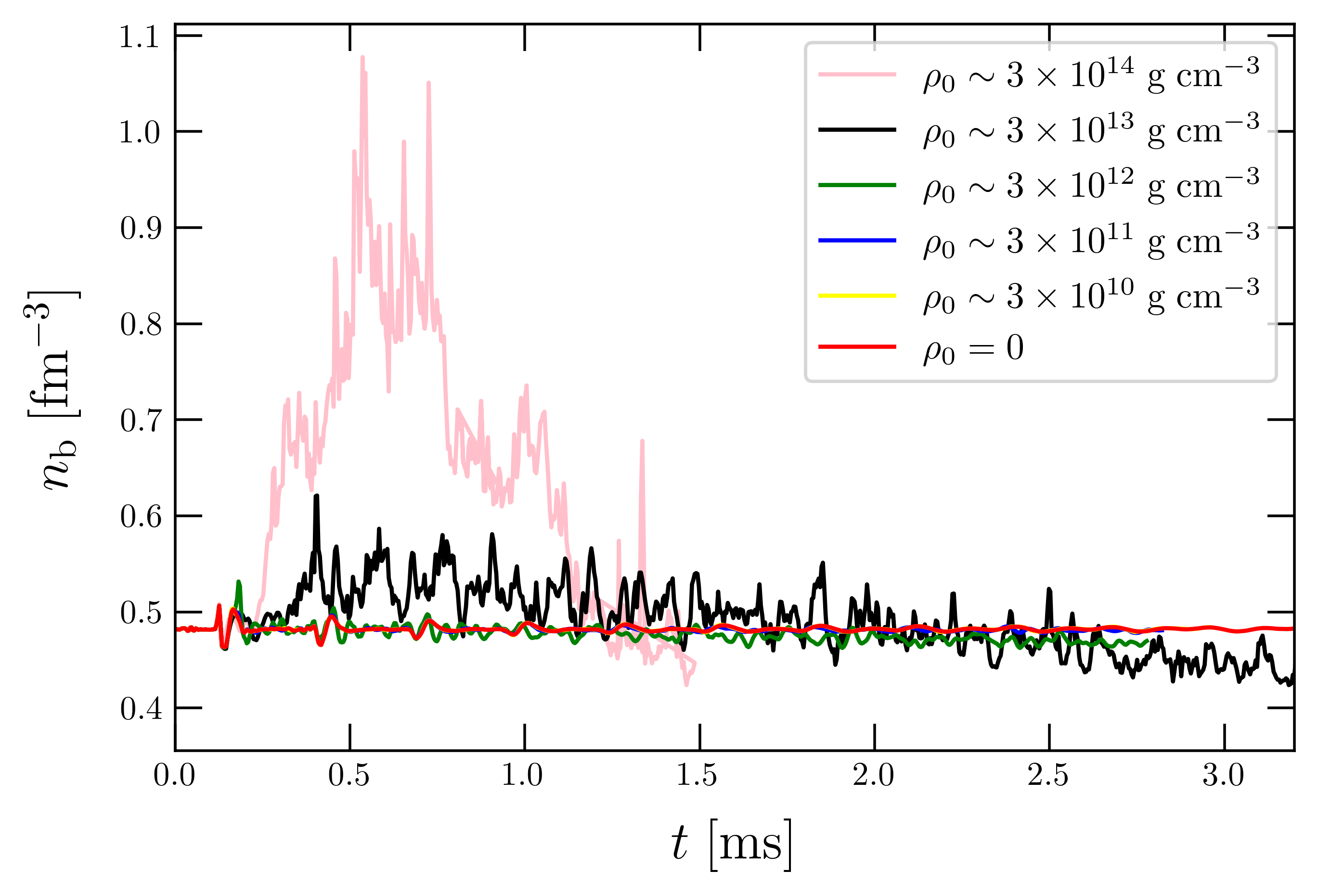}
\caption{\label{fig:rho_comparison} Central baryonic number density for an isolated, non-rotating strange star with different values of $\rho_0$. It is normalized to the nuclear saturation density $n_\mathrm{nucl} = 0.16 \; \rm{fm^{-3}}$. The initial value $\rho_0 \sim 3 \times 10^{14} \, \rm{g/cm^3}$ gives rise to serious oscillations which tend nicely to disappear as $\rho_0$ gets smaller.}
\end{figure}

For the TOV tests, we adopted the $T=0$ slice of the strange star EOS with a $\Gamma$-law to account for temperature evolution. Moreover, we work within the \emph{Cowling} approximation, i.e., the spacetime is fixed and unaffected by the hydrodynamics of the star. We briefly report the most significant results.

At first, we introduced a cut-off parameter $\rho_0$ to allow the code to distinguish between finite-pressure regions (with $\rho > \rho_0$) and the surface/floor ($\rho < \rho_0$, with $p \sim 0$). $\rho_0$ is a numerical threshold that might have served to artificially smooth the discontinuity at the surface. Despite that, we observe that such a cut-off parameter is not necessary. This is clarified in Fig. \ref{fig:rho_comparison}, showing the central baryonic density of a single strange star for different values of $\rho_0$.

Numerical perturbations that exist in any code yield oscillations of the central density in dynamical time scale even for TOV solutions. For strange stars, we observe these oscillations are significantly reduced, becoming less and less pronounced as $\rho_0$ is decreased. Eventually, when $\rho_0$ vanishes, we obtain the most stable configuration. This demonstrates our code's ability to resolve the surface of the strange quark matter stars without the need for any artificial smoothing of the surface. Given this evidence, we have set $\rho_0 = 0$ in all our simulations.

\begin{figure*}
\includegraphics[width=0.28\textwidth]{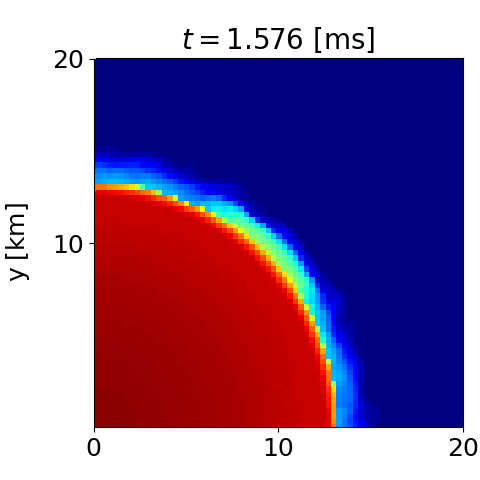}
\includegraphics[width=0.28\textwidth]{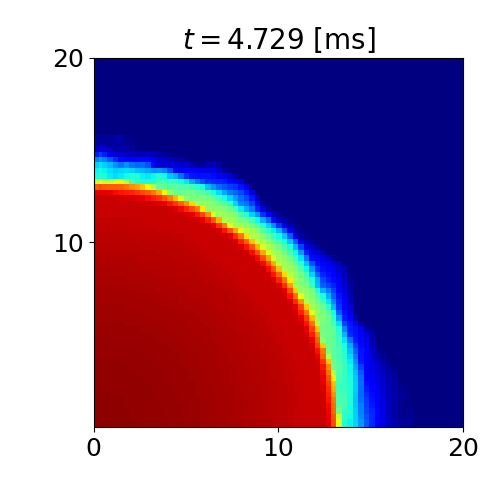}
\includegraphics[width=0.28\textwidth]{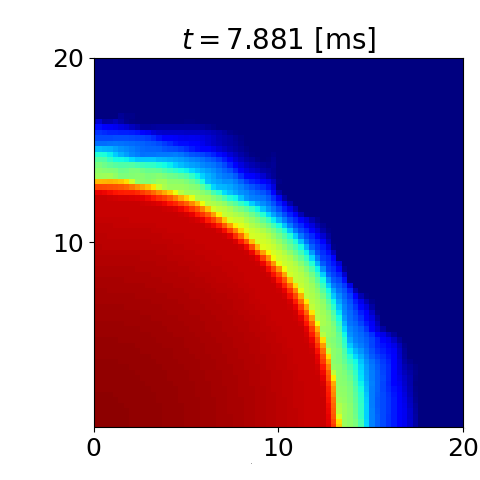}
\includegraphics[width=0.091\textwidth]{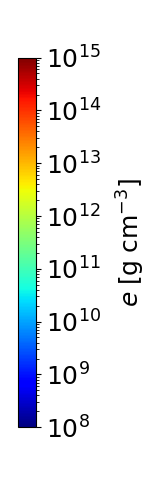}
\includegraphics[width=0.28\textwidth]{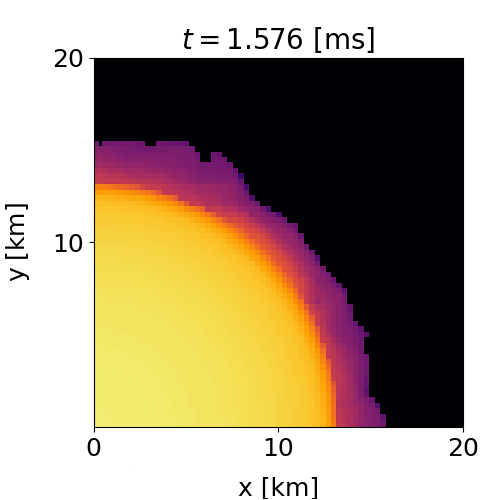}
\includegraphics[width=0.28\textwidth]{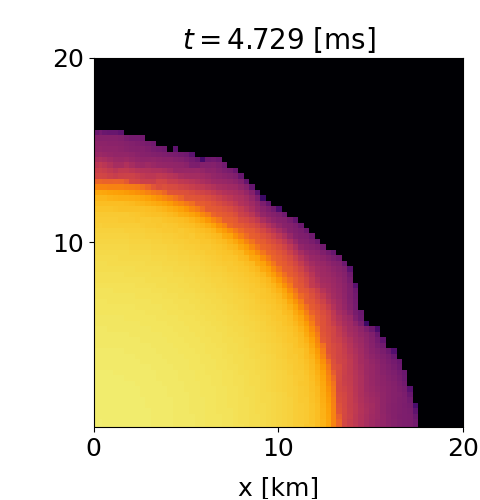}
\includegraphics[width=0.28\textwidth]{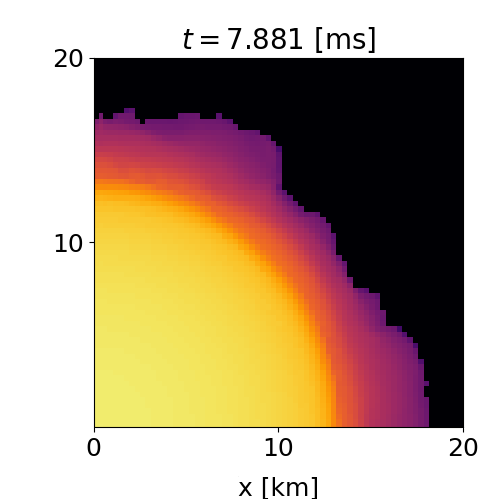}
\includegraphics[width=0.082\textwidth]{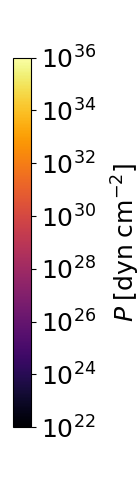}
\caption{\label{fig:e_P_no_rho}2D representative frames of the time evolution across $xy$ plane of energy density (upper panel) and pressure (lower panel) for an isolated, non-rotating strange star with no $\rho_0$. The stellar profile is very well-defined throughout the simulation time scale with a very thin layer of atmosphere at low $e$/$p$.}
\end{figure*}

This behavior is also stressed in Fig. \ref{fig:e_P_no_rho} where we show some characteristic equatorial slices of the energy density $e$ (upper panel) and the pressure $p$ (lower panel) at $\rho_0 = 0$. Even though the sequences are not so long (a few ms, but the behavior is the same for longer runs), the stellar profile is very well-defined with no significant spurious emission of material (which instead occurs with larger $\rho_0$). Rather, a thin layer of atmosphere surrounds the bare strange star.

To mitigate the numerical oscillations and to achieve good mass conservation, we also tested the effect of the implementation of the positive preserving limiter, sometimes referred to as \emph{pplim}. It usually allows for lower dissipation of mass (\cite{Radice:2013xpa}) at the cost of a slightly more widespread atmosphere. For such tests, we considered both single strange stars and representative binaries. 

The positivity preserving limiter is a hybrid scheme that ensures positive physical quantities (like the density $\rho$) remain such during the whole evolution of the system. We refer to \cite{Radice:2013xpa} for a comprehensive discussion, but the basic idea is hybridizing the computation of the flux from two sources: the first-order Lax-Friedrichs scheme and a high-order scheme. In particular, for a one-dimensional conservation law $\partial u / \partial t + \partial f(u) / \partial x = 0$, the discretization of the form:

\begin{equation}
    \label{eq:discretization_pplim}
    \frac{u_i^{n+1}-u_i^n}{\Lambda^0} = \frac{f_{i-1/2}-f_{i+1/2}}{\Lambda^1}
\end{equation}

can be optimized by choosing the flux term:

\begin{equation}
    \label{eq:flux_term_pplim}
    f_{i+1/2} = \theta f_{i+1/2}^{\mathrm{HO}} + (1-\theta) f_{i+1/2}^{\mathrm{LF}} \, ,
\end{equation}

where the apices HO and LF refer to the afore-mentioned schemes for the flux, and $\theta \in [0, 1]$ defines the weight of each scheme. In the bulk, i.e., far from the vacuum where the flow is smooth, usually $\theta = 1$, whereas in regions close to the vacuum where shocks may arise, $0 \leq \theta < 1$ is chosen so that the positivity is guaranteed.

\begin{figure*}
\includegraphics[width=0.49\textwidth]{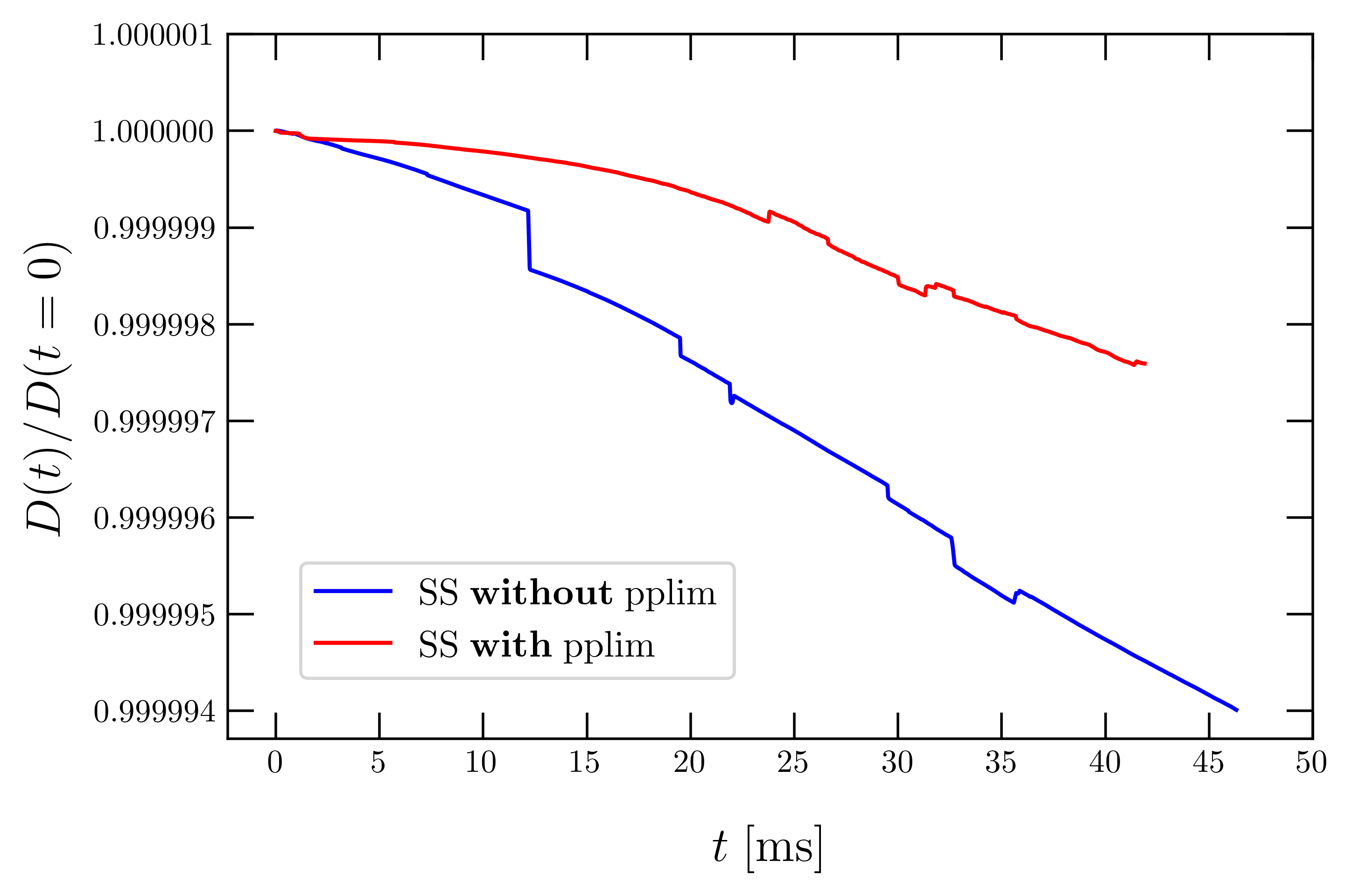}
\includegraphics[width=0.49\textwidth]{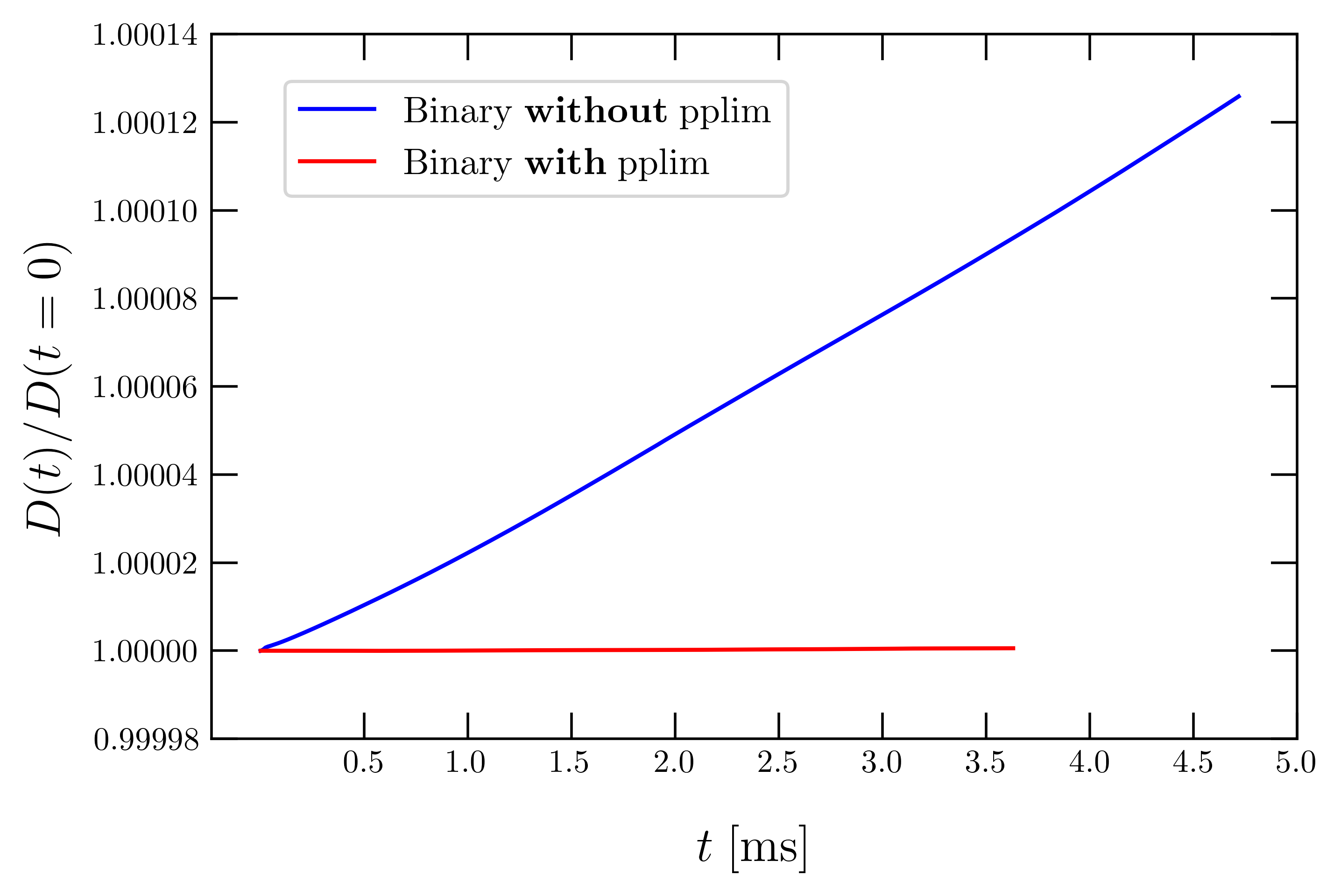}
\caption{\label{fig:pplim_D(t)} Normalized conserved density $D(t)/D(t=0)$ for an isolated strange star (left) and a binary system made of two strange stars (right) with and without the positivity preserving limiter. Numerical fluctuations are significantly smaller if the positivity preserving limiter is implemented.}
\end{figure*}

In Fig. \ref{fig:pplim_D(t)} we report the most important difference of taking into account the positivity preserving limiter or not. It shows the normalized conserved density $D(t)/D(t=0)$. The quantity $D = \rho W$ ($W$ is the relativistic Lorentz factor) is directly evolved and should in principle be constant. Numerical fluctuations may cause some loss of mass, but we observe that the inclusion of the positivity preserving limiter helps to conserve the rest-mass as demonstrated in Fig. \ref{fig:pplim_D(t)}. Interestingly, few ms are sufficient to appreciate how the effect of its implementation is even more relevant for the binary.

\begin{figure} [!ht]
\includegraphics[width=\columnwidth]{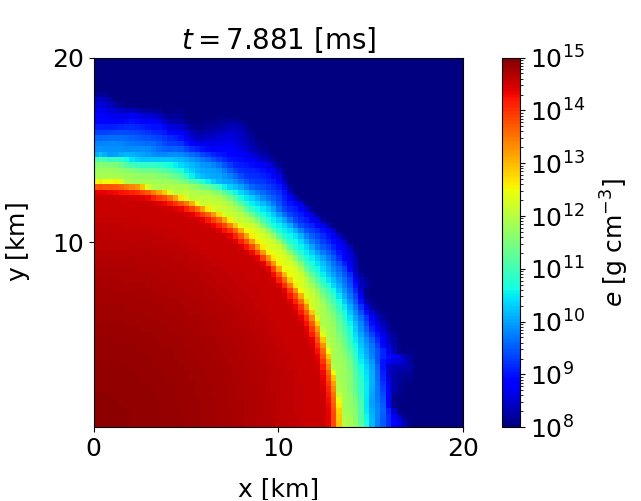}
\includegraphics[width=\columnwidth]{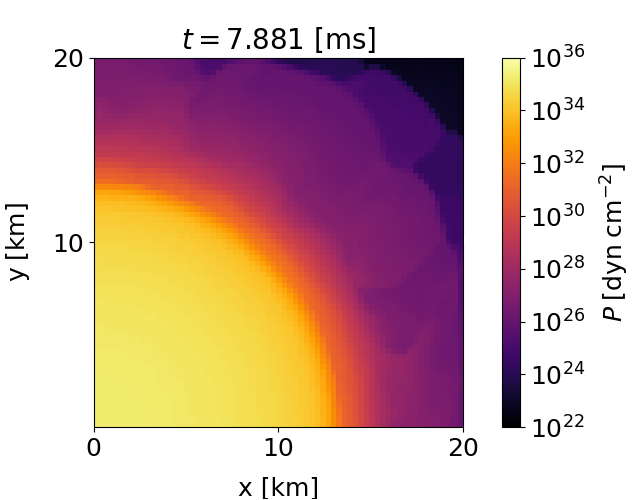}
\caption{\label{fig:pplim_rho_and_press}2D frames of energy density (top) and pressure (bottom) for bare, isolated, non-rotating strange star with $\rho_0 = 0$ when the positivity preserving limiter is implemented. Little material is spread out around the star, but the effect is negligible.}
\end{figure}

We finally show in Fig. \ref{fig:pplim_rho_and_press} a couple of frames, one for $e$ and the other one for $p$, where the positivity preserving limiter has been included. While the code spread the surface over a few more cells than Fig. \ref{fig:e_P_no_rho}, this does not yield a spurious secular evolution or artificial mass loss. Given that our numerical schemes ensure to capture the surface sharply (Figs. \ref{fig:rho_comparison}-\ref{fig:pplim_rho_and_press}) while also guaranteeing control over numerical fluctuations (Fig. \ref{fig:pplim_D(t)}), we employ the positivity preserving limiter in simulations of strange star mergers.

\clearpage

\bibliographystyle{apsrev4-1}
\bibliography{bibliography}

\end{document}